\documentclass[acmsmall,screen,authorversion]{acmart}

\usepackage{textcomp} % TM
\usepackage{subcaption} % subfigure
\usepackage{xcolor} 
\usepackage[ruled]{algorithm2e}
\newcommand{\revone}[1]{{#1}}

\acmPrice{15.00}

\copyrightyear{2022}
\acmYear{2022}
\setcopyright{acmlicensed}\acmConference[i3d]{I3D}{December xy--yz, 202X}{Somewhere, USA}
\acmBooktitle{I3D}
\acmPrice{15.00}
\acmDOI{10.1145/3478512.3488608}
\acmISBN{978-1-4503-9073-6/21/12}

\setcitestyle{square}

\begin{document}

% from the email - ACM Publishing License Text, Bibstrip, and Form - Proceedings of the ACM on Computer Graphics and Interactive Techniques 175
\setcopyright{acmlicensed}
\acmJournal{PACMCGIT}
\acmYear{2023} \acmVolume{6} \acmNumber{1} \acmArticle{} \acmMonth{5} \acmPrice{15.00}\acmDOI{10.1145/3585503}

% Title. 
% If your title is long, consider \title[short title]{full title} - "short title" will be used for running heads.
\title{ Subspace Culling for Ray--Box Intersection }

% Authors.
\author{Atsushi Yoshimura}
\affiliation{
  \institution{Advanced Micro Devices, Inc.}
  \country{Japan}
  }
\email{Atsushi.Yoshimura@amd.com}

\author{Takahiro Harada}
\affiliation{
  \institution{Advanced Micro Devices, Inc.}
  \country{USA}
  }
\email{Takahiro.Harada@amd.com}

% \author{Brittany Rowland-Smith}
% \affiliation{%
%   \institution{St. Olaf College}}
% \email{br-s@gmail.com}

% \author{Nicholas Badeeri}
% \affiliation{%
%   \institution{MathWorks, Inc.}}
% \email{badeeri@mathworks.com}

% \author{Andrew Joseph Foyt}
% \affiliation{%
%   \department{College of Engineering}
%   \institution{University of Houston}}
% \email{foyt_aj@uh.edu}

% This command defines the author string for running heads.
% \renewcommand{\shortauthors}{DeJohnette, Rowland-Smith, Badeeri, and Foyt}
% \renewcommand{\shortauthors}{Tagliasacchi et al.}

% abstract
\begin{abstract}
Ray tracing is an essential operation for realistic image synthesis. 
The acceleration of ray tracing has been studied for a long period of time because algorithms such as light transport simulations require a large amount of ray tracing. 
One of the major approaches to accelerate the intersections is to use bounding volumes for early pruning for primitives in the volume. The axis-aligned bounding box is a popular bounding volume for ray tracing because of its simplicity and efficiency.
However, the conservative bounding volume may produce extra empty space in addition to its content. Especially, primitives that are thin and diagonal to the axis give false-positive hits on the box volume due to the extra space.
Although more complex bounding volumes such as oriented bounding boxes may reduce more false-positive hits, they are computationally expensive.
In this paper, we propose a novel culling approach to reduce false-positive hits for the bounding box by embedding a binary voxel data structure to the volume.
As a ray is represented as a conservative voxel volume as well in our approach, the ray--voxel intersection is cheaply done by bitwise AND operations. 
Our method is applicable to hierarchical data structures such as bounding volume hierarchy (BVH). It reduces false-positive hits due to the ray--box test and reduces the number of intersections during the traversal of BVH in ray tracing.
We evaluate the reduction of intersections with several scenes and show the possibility of performance improvement despite the culling overhead. We also introduce a compression approach with a lookup table for our voxel data. We show that our compressed voxel data achieves significant false-positive reductions with a small amount of memory.
\end{abstract}

%CCS


%keywords

\begin{teaserfigure}
\begin{tabular}{@{\hspace{0\linewidth}}c@{\hspace{0.005\linewidth}}c@{\hspace{0.005\linewidth}}c@{\hspace{0.002\linewidth}}} 
\begin{subfigure}[b]{0.32\textwidth}
    \includegraphics[width=1\linewidth, trim=100 0 100 0, clip]{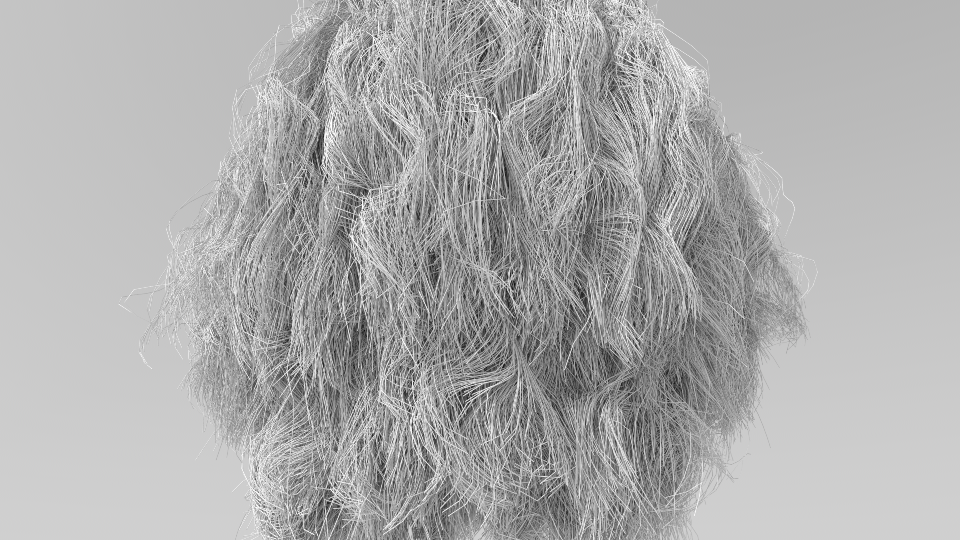} 
    \caption{ Curly Hair scene }
\end{subfigure}
\begin{subfigure}[b]{0.32\textwidth}
    \includegraphics[width=1\linewidth, trim=200 0 200 0, clip]{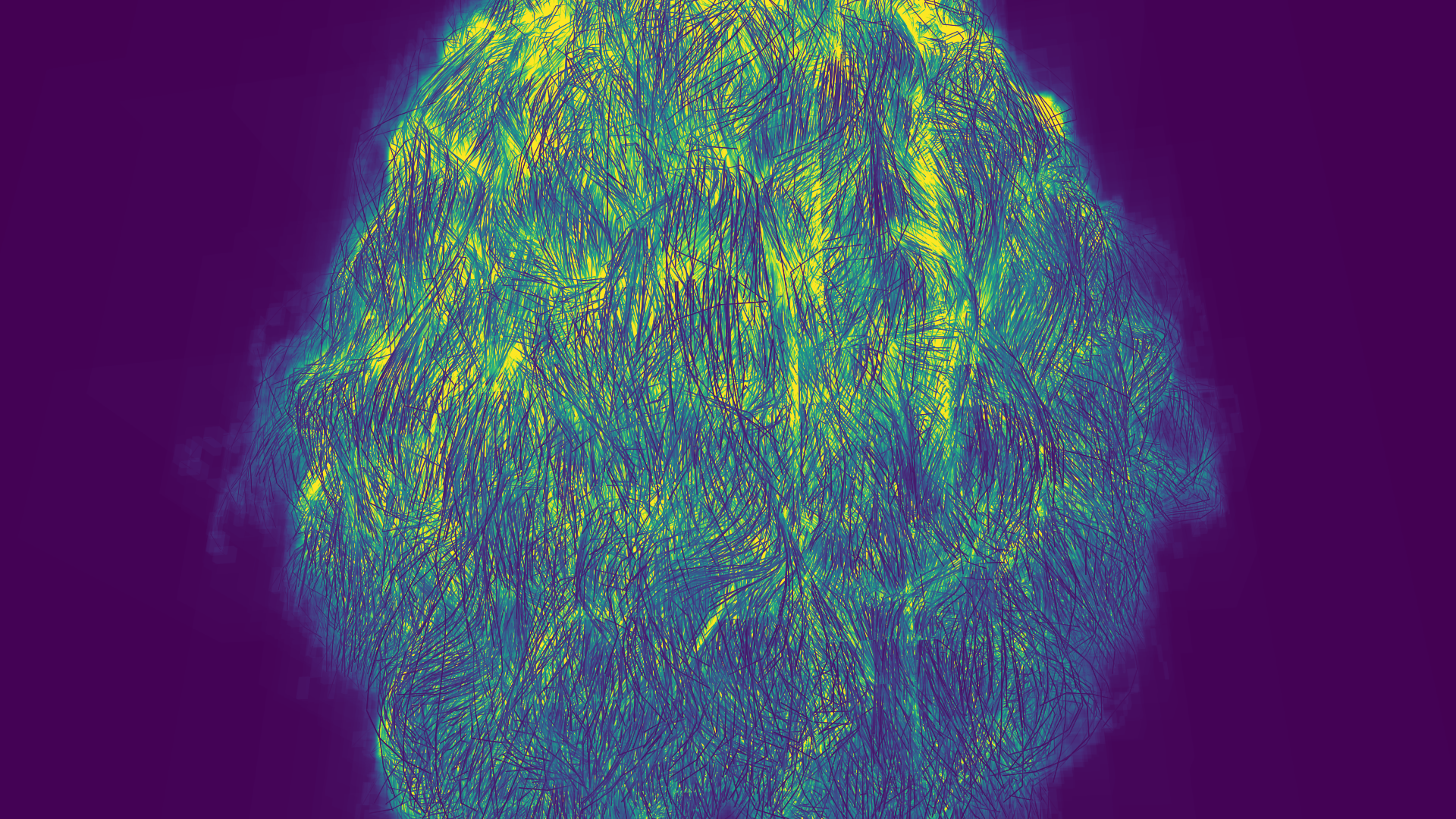} 
    \caption{ without our culling }
\end{subfigure}
&
\begin{subfigure}[b]{0.32\textwidth}
    \includegraphics[width=1\linewidth, trim=200 0 200 0, clip]{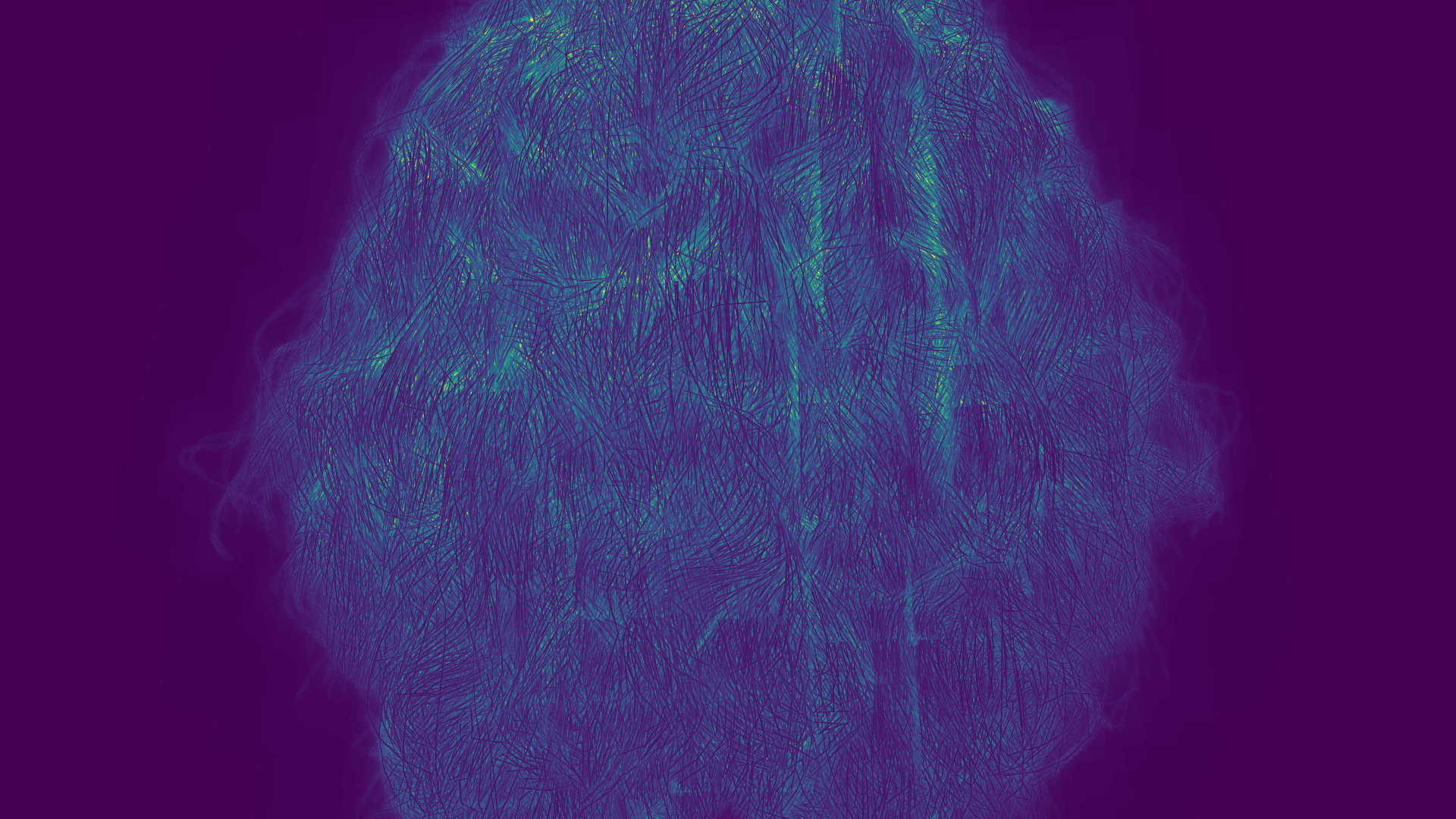} 
    \caption{ with our culling }
\end{subfigure}
&
\begin{subfigure}[b]{1.0\textwidth}
    \includegraphics[height=0.227\linewidth]{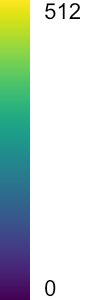} 
    \caption{  }
\end{subfigure}
\end{tabular}
\caption{Visualization of the number of intersections on the primary rays with and without our culling, where the number of bits on the voxel data structure for each AABB is 216. The intersection between a ray and the voxels is accelerated by our look-up table-based approach.  (a): The scene consists of 12.1 million triangles and almost all of the triangle is thin and tilted. (b) and (c): The intersection count of triangles and internal nodes is mapped to the color with a range of $0$ - $512$. 
Our method reduces the number of intersections by 37.5\% and decreases the rendering time by 13.1\% for this scene.
}
\label{teaser}
\end{teaserfigure}
\maketitle
\section{Introduction}
Ray tracing is a method widely used for various applications to find intersections with scene primitives. Algorithms such as path tracing rely heavily on ray tracing for photo-realistic image rendering. However, a vast number of ray queries consume significant computational resources. Several data structures and algorithms such as grid traversal \cite{fastvoxel}, octrees \cite{octrees}, and the bounding volume hierarchy \cite{HierarchicalGeometric} have been introduced and studied in order to accelerate ray tracing. These approaches use one or more kinds of bounding volumes to reduce the number of intersections by skipping the primitive tests that are assigned to the volume if it does not hit. The axis-aligned bounding box (AABB) is one of the most popular bounding volumes because of its simplicity and efficiency in ray traversal and data structure building. 
However, AABB may not be able to tightly fit the primitives such as, hair and fur, that are thin and diagonal to the axis.
More spatially flexible bounding volumes, such as the oriented bounding boxes (OBB), can be used \cite{HairAndFur}. Although its culling efficiency is better, the computational cost and memory consumption are high. Instead, we propose a novel culling technique subsequent to the AABB-ray test to reduce the number of false-positive intersections produced by AABB.

Our method effectively reduces the total number of intersections by embedding a binary voxel data structure for each node in the bounding volume hierarchy (BVH), as shown in Fig.~\ref{teaser}. A ray is represented as a binary voxel volume with the same resolution. The intersection is done by bitwise AND operations. We also introduce using a lookup table (LUT) for voxel data compression, which reduces the memory size without sacrificing a lot of culling efficiency and traversal performance. 

\section{Related Work}
A voxel data structure can be used for ray tracing acceleration by skipping cells that are not along the rays.
Variants of a digital differential analyzer (DDA) was proposed for ray tracing \cite{3DDDA, fastvoxel}.
Fine voxels can fit primitives tightly even if they are concave; however, the data structure requires iterative algorithms to find all intersected cells with a ray, and the number of iterations is proportional to the voxel resolutions.

Intersections between a ray and a chunk of binary voxels can be accelerated by blockwise packing \cite{GPUPro}. Voxels along a ray are also packed as the same representation as the scene voxels, and the intersection is efficiently done by bitwise AND operations.

The bounding volume hierarchy \cite{HierarchicalGeometric, BVHStar} effectively reduces the number of intersections by hierarchical culling with bounding volume such as AABB. Although AABB is the most popular choice for the bounding volume, it suffers from false positive intersections due to non-tightly fit primitives. OBB can be used for further reduction of the false positives \cite{HairAndFur}; however, the computational overhead is higher, and larger memory space is needed.
% however, it has to pay computational overhead and extra memory space. 
Tessellating primitives into finer granularity also alleviates the false positives \cite{SBVH1, SBVH2}, but the drawback is memory consumption due to the deeper bounding volume hierarchy.
Molenaar and Eisemann proposed another method using a proxy geometry represented by a hierarchical voxel data structure to optimize out-of-core rendering \cite{ConservativeRayBatching}. Our approach is similar to theirs but our method has a smaller overhead which makes it possible to use at each step in the traversal, and it pays off for in-core rendering.
%\revone{Another approach to reducing the number of intersections is to use a proxy geometry represented by a hierarchical voxel data structure for culling before traversing certain levels of BVH \cite{ConservativeRayBatching}. 
%The proxy-based culling can reduce the number of rays that invoke locally expensive operations such as data transfer and BVH construction. 
%In contrast, it is suitable for a case where the BVH traversal cost is uniform as our culling occurs in each traversal step.}

We propose a novel culling that achieves a coexistence of culling efficiency, cheap computation, and low memory cost.
Our culling leverages voxels' tightness and intersection efficiency to compensate for the AABB weakness. The voxels can be compressed with LUT-based compression to minimize the memory cost.

% \takahiro{[Give a high level explanation here? It starts with the detail.]} 
\section{Our culling approach}
\begin{figure}
\centering
\includegraphics[width=0.7\linewidth,trim=0 0 0 0, clip]{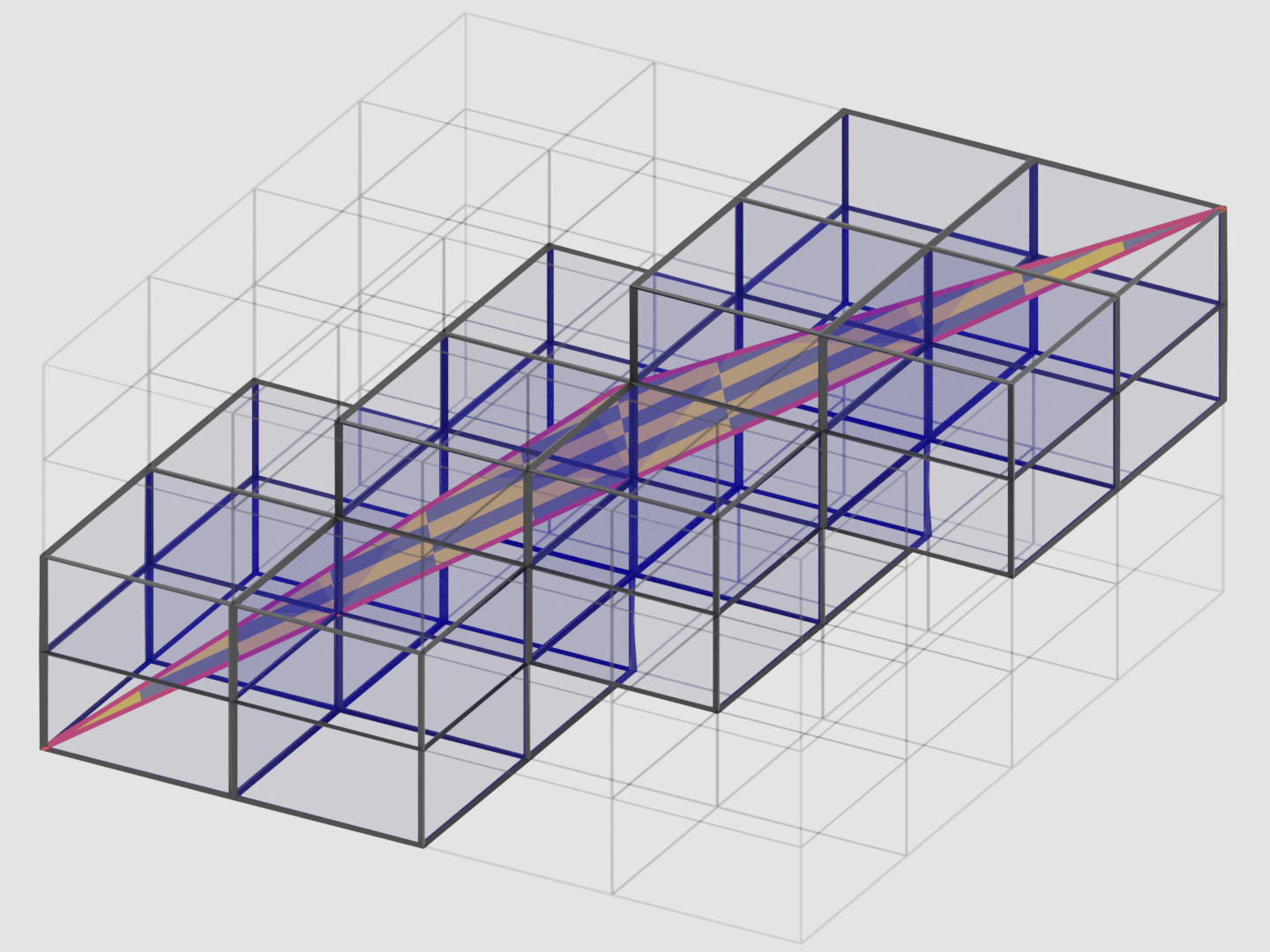} % trim: left, bottom, right and top
\caption{ A triangle bounded more tightly than AABB by $4 \times 4 \times 4 $ voxels. Only 22 of 64 voxels are occupied by the triangle. }
\label{fig:ourvoxels}
\end{figure}

\begin{figure}
\centering
\includegraphics[width=1.0\linewidth]{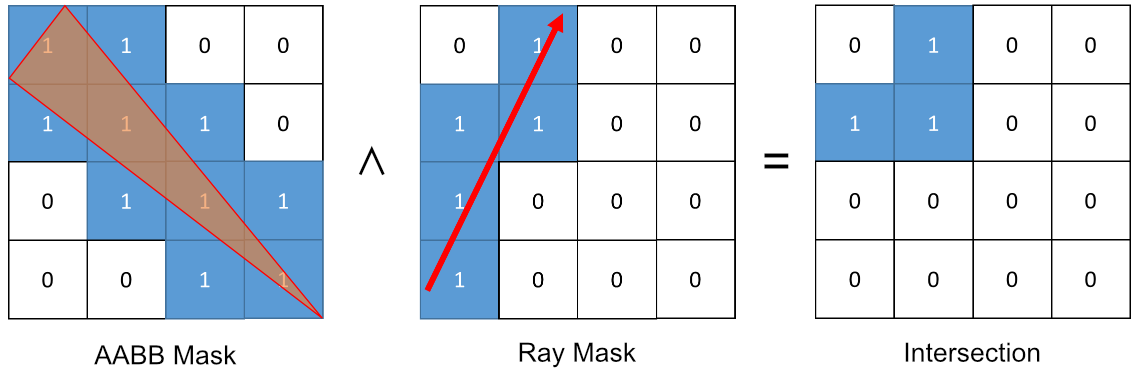} % trim: left, bottom, right and top
\caption{ An example of Ray-AABB intersection by AABB mask and ray mask by a bitwise AND operation. If all bits in the result are zero, there is no intersection. }
\label{fig:isect}
\end{figure}

AABB is a simple and efficient bounding volume; however, primitives such as a long, thin, and diagonal triangle in the volume may not be tightly bounded.
Our approach compensates for this loose bound due to AABB by a uniform grid and storing occupancy of a triangle as shown in Fig.~\ref{fig:ourvoxels}.
Since the voxel data structure can represent empty space for each cell individually, it can more tightly enclose these kinds of primitives.
We build a uniform grid data structure representing spatial occupancy as binary for each AABB in a BVH, which we call an object mask. 
In order to determine the voxel pattern along a ray in the same grid as the object mask, AABB-ray intersections are needed, such as the DDA-based methods \cite{3DDDA, fastvoxel}.
However, such iterative algorithms for each AABB during BVH traversal may be impractical because the grid traversal takes linear time complexity with respect to the grid resolution.
% A traversal through the uniform grid along a ray can be done by an iterative algorithm \cite{fastvoxel}. 
Instead, we build a LUT for the overlapped voxels with a ray in the grid. Accordingly, the voxel pattern is obtained by looking up the LUT. We call the voxel pattern for a ray mask. 
The intersection between a ray mask and an object mask can be executed by bitwise AND operations, as shown in Fig.~\ref{fig:isect}. 
Thus, our culling in subspace can reduce false positives due to AABB by simple arithmetic operations.

% We do an extra test to reduce false-positive hits after Ray--AABB intersection is detected.
% We use a small binary voxel mask to represent the spatial occupancy of primitives in the AABB. A ray is represented in the same manner as the AABB mask. We check the intersection by logical "and" as shown in Fig.~\ref{fig:isect}.
First, we describe our method with a simple case — triangles in an AABB in Sect. 3.1, then extend it to BVH, a hierarchical case, in Sect. 3.2.

\subsection{Subspace Culling for Triangle-Ray Intersection} 

\subsubsection{Mask Creation}
We create an object mask and a ray mask to test the overlap. In order to build an object mask in an AABB containing a single triangle (or triangles), we conservatively voxelize the triangles \cite{FastVoxelization}. A chunk of binary occupancy can be densely packed as a bit sequence.
% an integer variable if the voxel resolution is below $4^3$. % I stopped mentioning integers.

We precompute ray masks as a LUT. We use two discretized positions of AABB-ray intersections as the key of the LUT; thus, the table has six dimensions. The hit locations by AABB-ray intersection are linearly mapped to a unit cube then the indices are determined as shown in Algorithm~\ref{alg:lookup} to reuse the LUT for all AABB.
The two positions of AABB-ray intersections can be anywhere in the AABB in cases where the ray origin or the end of the ray segment is inside the volume, as long as the ray mask LUT includes masks corresponding to all combinations of the two positions.
We use AABB-AABB Sweep intersection test \cite{RealTimeCollision} to build the LUT because the sweep represents all possible rays. Although the lookup-based ray mask is conservative, it may produce extra false-positive intersections, which we evaluate in Sect. 4.2.

\begin{algorithm}[!t]
\caption{ Lookup ray mask \textbf{Inputs:} \textit{lower, upper}: Minimum and maximum value of the AABB.
\textit{$p_0$, $p_1$}: The hit locations by AABB-ray intersection.
\textit{R}: The mask resolution.
\textit{rayMasks}: The LUT for a ray mask. Discretized begin and end location of the ray map to its ray mask. }
\label{alg:lookup}
\SetKwProg{Fn}{function}{}{end}
\SetKwFunction{lookupRayMask}{lookupRayMask}
\Fn{\lookupRayMask{ lower, upper, $p_0$, $p_1$, R, rayMasks }}
{
    {$extent$ $\gets$ $ upper - lower $ }\;
    {$beg$ $\gets$ $ \min( \lfloor \frac{ R }{extent} (p_0 - lower) \rfloor, R - 1) $ }\;
    {$end$ $\gets$ $ \min( \lfloor \frac{ R }{extent} (p_1 - lower) \rfloor, R - 1) $ }\;
    \Return $rayMasks_{( beg, end )}$\;
}
\end{algorithm}

\subsubsection{ Culling by Ray Mask and Object Mask }
Once the ray mask LUT and object masks are created, the culling is straightforward. First, we compute the two intersection points of the ray on the AABB, which are used to fetch a ray mask from the LUT. Then, we take bitwise AND between the ray mask and the object mask. If all the bits are zero, there is no intersection between the ray and the AABB. 
Therefore ray-triangle intersections can be culled by the masks further after AABB culling.

\subsection{ Hierarchical Subspace Culling for BVH }
Our method can be trivially applied to every node in BVH because each node can be considered a node containing all of the triangles in descendant nodes. However, calculating the occupancy with all primitives in a node is computationally expensive, especially for nodes in an upper level on a BVH. 
Thus, we propose an approximation to accelerate object mask creation dedicated to BVH. 
Additionally, object masks for each AABB can consume large memory. We also propose to use LUT to compress object masks in a BVH to reduce the memory overhead.
% ↑ with considering the relation with the previous section.

% Each BVH node often uses AABB. Hierarchical AABB can be tightly bounded by an AABB by finding minimum and maximum boundaries just for its direct children. 
% The tightest object mask requires checking the occupancy with all primitives in the volume, however this is computationally expensive. Thus, we propose a few methods to approximate the occupancy by its object mask instead of all primitives in the volume. Also, we show that object masks can be compressed by LUT to reduce the memory overhead in BVH.

\subsubsection{Hierarchical Object Mask Creation}
\begin{figure}
    \centering
    \begin{subfigure}[b]{0.31\textwidth}
    \includegraphics[width=1\linewidth]{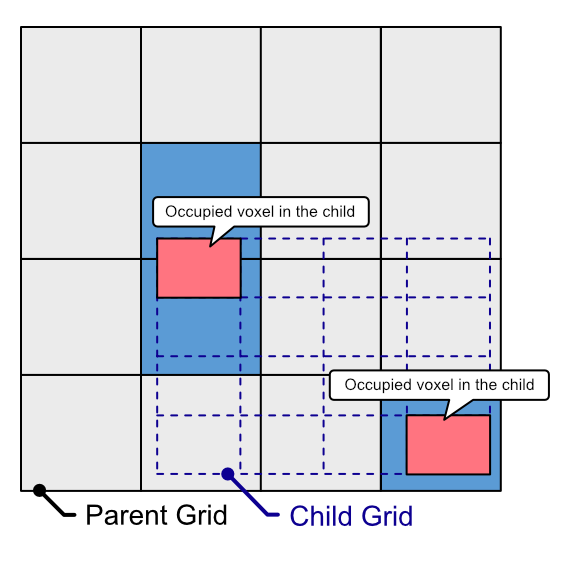} 
    \caption{ Voxel filling in the parent }
    \label{fig:hierarchical:a}
    \end{subfigure}
    \hfill
    \begin{subfigure}[b]{0.31\textwidth}
    \includegraphics[width=1\linewidth]{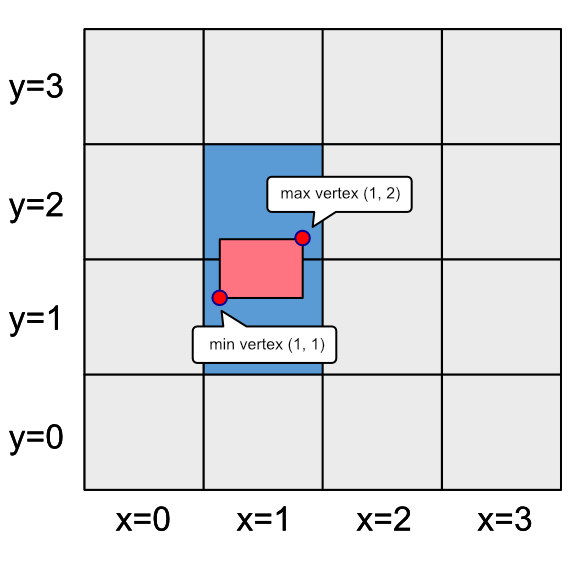}
    \caption{ Two voxels are overlapped }
    \label{fig:hierarchical:b}
    \end{subfigure}
    \hfill
    \begin{subfigure}[b]{0.31\textwidth}
    \includegraphics[width=1\linewidth]{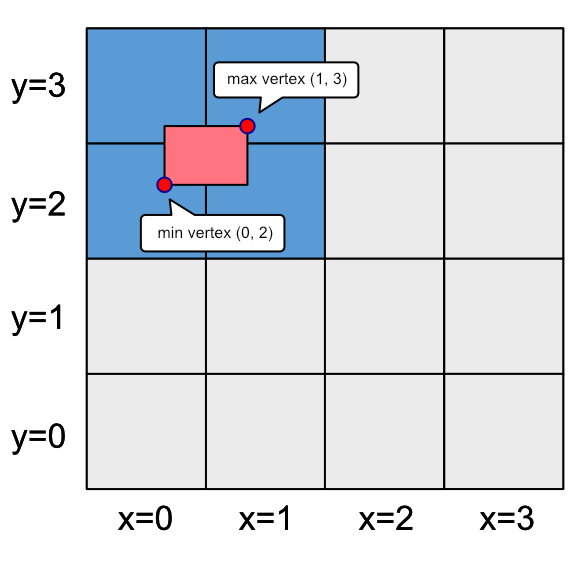}
    \caption{ Four voxels are overlapped }
    \label{fig:hierarchical:c}
    \end{subfigure}
    \caption{The parent voxels are conservatively filled by their occupied voxels in children. Occupied voxels in the parent and occupied voxels in the child are illustrated as blue and red cells, respectively. (a): Three voxels in the parent are filled by two voxels in the child. (b) and (c): Two box-shaped filling patterns in the parent by a voxel in a child. A child voxel fills one to eight parent voxels since a child voxel size is smaller than or equal to the parent. }
    \label{fig:hierarchical}
\end{figure}

As a node may contain a large number of primitives due to the hierarchical representation, voxelizing all the primitives to compute the occupancy is redundant and computationally expensive. 
Instead, we introduce the use of an object mask on a node as an approximation of the occupancy of its children. 
\revone{An occupied voxel in the child conservatively fills its parent voxels as shown Fig.~\ref{fig:hierarchical:a}.
Fig.~\ref{fig:hierarchical:b} and Fig.~\ref{fig:hierarchical:c} show two filling patterns in the parent from a child voxel. A child voxel's min and max vertices are projected to the parent grid. Their indices determine the box-shape filling pattern in the parent node. Because the pattern is finite, 
we precompute a six-dimensional table for the box-shaped pattern. The table maps the projected indices of the vertices to their filling pattern. 
Each occupied child voxel is mapped to a pattern and they are combined by bitwise OR operations to get the conservative voxels in the parent. 
Algorithm~\ref{alg:fillChildMask} shows an object mask creation from the approximated occupancy of its children.}
% A child voxel fills one to eight voxels since a child voxel is smaller than or equal to the parent voxel.
% First, grid vertices of an object mask on a child are projected into indices of the parent grid of its object mask.
% The parent mask's voxels that correspond to each occupied voxel in the child's mask are set using the projected indices.
% Since a voxel in the child mask can overlap multiple voxels in the parent, these overlapped voxels are calculated by the lower and upper vertices of the voxel in the child. 
% The overlapped voxel pattern on the parent mask is applied to the parent object mask by bitwise OR operations. 
% Additionally, as the voxel pattern is always box-shaped, the voxel pattern creation out of the two voxels can be replaced with fetching from a precomputed pattern table, which simplifies the occupancy propagation from the child to its parent.
% Algorithm~\ref{alg:fillChildMask} shows an object mask creation from the approximated occupancy of its children.
As the approximation may produce extra false positives, we use a parameter $L$ to control the strength of approximation, where we compute object masks from the node $L$ levels below to take tighter occupancy. As $L$ increases, the computational cost increases as we need to propagate object masks and primitives to many parents. $L=\infty$ is equivalent to the tightest object mask built from all primitives below the node \revone{while $L=1$ limits a node to take the occupancy only from its direct children.}
% \takahiro{[Just refering Algorithm here isn't enough. Need to explain what it does in text even if it's short. (for example transform the aabb of a child node to the node we are computing the mask, then go through all the voxels and mark the occupied voxel in the child etc) then we can refer algorithm for the detail. Also no explanation about BoxMasks?]} 

\subsubsection{Object Mask Compression with LUT}
The memory size of an object mask is proportional to the cube of the resolution $R$. For instance, we need to store a 64-bit value for an AABB when we use $R=4$.
In order to reduce the memory footprint of object masks, we propose the use of a LUT to compress them. Each mask data is replaced by an index of the LUT, which we call a compression LUT.
The tightness of the compressed masks depends on a set of object masks we store in the compression LUT.
We borrow the idea of surface area heuristic (SAH) \cite{SAH1, SAH2} to define the optimal LUT. In SAH, the cost of BVH is described as follows:

\begin{equation} \label{math:sah}
SAH( N) =\frac{1}{SA( N)}\left( C_{T}\sum _{N_{i}} SA( N_{i}) +C_{I}\sum _{N_{l}} SA( N_{l}) \cdot |N_{l} |\right),
\end{equation}
where $N$ is the root node, $SA(N)$ is the surface area of the bounding box on the node, $N_i$ and $N_l$ are inner nodes and leaf nodes, $C_T$ and $C_I$ are intersection cost of an inner node and a primitive, $|N_{l} |$ is the number of primitives in the leaf.

We assume the number of occupancy in the object mask is proportional to the probability of the intersecting ray. 
Then the SAH can be extended to take our culling into account as follows:

\begin{equation} \label{math:sahculling}
SAH_{masked}( N) =\frac{1}{SA( N)}\left( C_{T}\sum _{N_{i}} SA( N_{i})\frac{O_{m}( N_{i})}{R^{3}} +C_{I}\sum _{N_{l}} SA( N_{l})\frac{O_{m}( N_{l})}{R^{3}} \cdot |N_{l} |\right),
\end{equation}
where $O_{m}(x)$ is the number of occlusions in the object mask in a node $x$. As a compression LUT has a limited number of elements, the LUT that minimizes Eq.~\ref{math:sahculling} is the optimal table. 

However, it is prohibitively expensive to find the optimal LUT because of the vast search space. Thus, we first choose the object masks from the original masks for a compression LUT. We use $ SA(N)\frac{O_{m}( N)}{R^{3}} $ as the importance of its object mask for creating a sub-optimal LUT. We take the top important object masks as a compression LUT. Second, we find the conservative and the tightest object mask from the LUT for each mask that is not found in the LUT.

\revone{Note that the intersection of the LUT for ray mask and compression LUT can be obtained by a precomputed two-dimensional bit table. The table key consists of an index of the ray mask and an index of the compression LUT. The table values are the intersection result represented as a bit.}
Accordingly, fetching a ray mask and an object mask from each LUT and bitwise AND operations can be replaced by fetching an intersection result as a bit from the precomputed bit table. This replacement can improve the performance of the intersection if the bit fetching is cheaper despite memory access to random locations in the large bit table.
% Note that the intersection of the LUT for ray mask and compression LUT can be precomputed as another LUT if the element count of ray mask LUT times the element count of compression LUT is small enough.
% This reduces the memory traffic of both the ray mask and object mask.

\newcommand*\BitOr{\oplus}

\begin{algorithm}[!t]
\caption{ An object mask creation from the approximated occupancy. The indices of the parent voxel grid are calculated individually per axis to reuse the result in the loop for each voxel of the child mask. \textbf{Inputs and Symbols:} \textit{mask}: AABB mask to update its occupancy. 
\textit{pLower, pUpper}: Minimum and maximum value of the parent AABB.
\textit{childMask}: Approximated occupancy of the child.
\textit{cLower, cUpper}: Minimum and maximum value of the child AABB.
\textit{R}: Mask resolution.
\textit{\revone{fillingPatternTable: A table maps minimum and maximum voxel indices in a parent grid to a box-shaped voxel pattern.
$\BitOr$: A bitwise OR.}}}
\label{alg:fillChildMask}
\SetKwProg{Fn}{function}{}{end}
\SetKwFunction{fillByApproximatedOccupancy}{fillByApproximatedOccupancy}
% \SetKwFunction{buildBoxMasks}{buildBoxMasks}
\Fn{\fillByApproximatedOccupancy{ mask, pLower, pUpper, childMask, cLower, cUpper, R, fillingPatternTable }}
{
    {$cExtent$ $\gets$ $cUpper - cLower$ }\;
    {$pExtent$ $\gets$ $pUpper - pLower$ }\;
    \For{$i \gets 0$ to $R$} 
    {
        {$border$ $\gets$ $ cLower + \frac{i}{R}cExtent $ }\;
        {$indexParent$ $\gets$ $ \lfloor R\frac{ border - pLower }{pExtent}\rfloor $ }\;
        {{$xs_{i}, ys_{i}, zs_{i}$} $\gets$ $ \min( indexParent, R - 1 )$}\;
    }
    \BlankLine
    \tcp{A Loop for each voxel of the child mask.}
    \For{$x \gets 0$ to $R$ - 1} {
    \For{$y \gets 0$ to $R$ - 1} {
    \For{$z \gets 0$ to $R$ - 1} {
        \If{ $childMask_{xyz} = 0$ }
        {
            \textbf{continue}
        }
        {$index_{min}$ $\gets$ {$xs_{x}, ys_{y}, zs_{z}$}}\;
        {$index_{max}$ $\gets$ {$xs_{x+1}, ys_{y+1}, zs_{z+1}$}}\;
        {$mask$ $\gets$ $ mask \BitOr fillingPatternTable_{ ( index_{min}, index_{max} ) } $ }\;
        
        % \textbf{TODO: This can be replaced by LUT. It is even faster. }
        % {$m$ $\gets$ 0}\;
        % {$m_{index0}$ $\gets$ 1}\;
        % \For{$dstx$ $\gets$ $index0_{x}$ to $index1_{x}$ } {
        %     { $m$ $\gets$ $ m \ll 1 $ }\;
        % }
        % \For{$dsty$ $\gets$ $index0_{y}$ to $index1_{y}$ } {
        %     { $m$ $\gets$ $ m \ll R $ }\;
        % }
        % \For{$dstz$ $\gets$ $index0_{z}$ to $index1_{z}$ } {
        %     { $m$ $\gets$ $ m \ll R^2 $ }\;
        % }
        % { $mask$ $\gets$ $ mask | m $ }\;
        % 
    }
    }
    }
}
\end{algorithm}

\subsubsection{Search Object Mask in the LUT }
Some object masks may not be in the compression LUT when the number of LUT elements is less than object masks in a BVH.
We need to select an alternative mask from compression LUT for such missing masks. The alternative mask needs to be conservative; otherwise, it causes invalid culling. The alternative mask is preferred to be a tighter one in the conservative masks to make Eq.~\ref{math:sahculling} smaller.
Accordingly, we extract masks that have all set bits of the missing object mask out of compression LUT and choose the mask with the smallest number of set bits. The brute-force algorithm can do this; however, we propose a search algorithm with which we can search for the best mask in table look-ups and a few arithmetic operations. The algorithm consists of conservative mask extraction and searching for the tightest one. The former is done using precomputed tables, and the latter is done by presorting the compression LUT by their set bit count.

% Even if the masks with higher scores (SAHs) are selected, there is no guranttee that all masks we use in the BVH are included in the list. 
% For those masks, we select a mask close to the one for the AABB. The selection needs to be conservative as we do not want to miss intersection. However, we do not want to use the mask where all the voxels are marked as filled. Therefore, the criteria of the conservative mask selection is all the voxels in the original mask is set and it should have as small number of occupied voxels excluding the occupied voxels in the original mask as possible.  
% We could use the brute-force algorithm (there is only one brute force algorithm thus the, not a I think) when the size of the LUT is small. However, we propose a search algorithm with which we can search for the best mask in a few table look ups and arithmetic operations.

\begin{figure}
\centering
\includegraphics[width=1\linewidth]{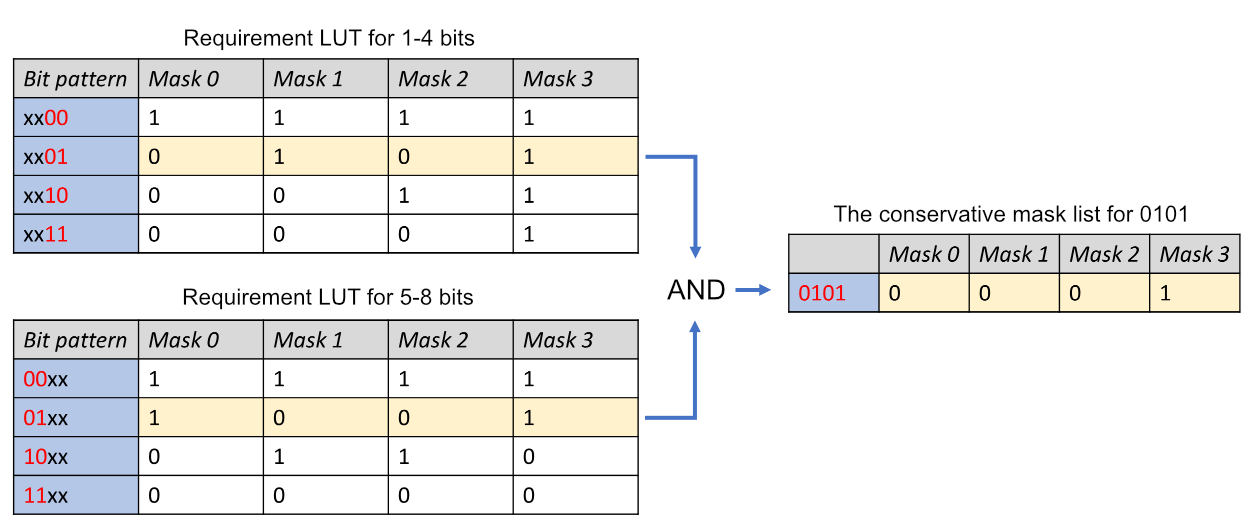} % trim: left, bottom, right and top
\caption{ An example of requirement LUTs with $b=2$. A bit pattern of an object mask is separated into $b$ bits elements, and a mask list that corresponds to its bit pattern from each requirement LUT is combined by bitwise AND operations to get the conservative mask list for the object mask to search. }
\label{fig:lutsearch}
\end{figure}

Since the set bits in the object mask indicate the requirements, we precompute the list of masks that satisfy the requirements in the compression LUT against each bit pattern. The list of the masks is stored as a bit sequence where each bit represents whether the mask corresponding to its bit location satisfies the requirements.
However, the bit pattern is $2^{R^3}$, which is impractically large to compute. In order to reduce the size of the precomputed table, we separate the bit pattern into smaller chunks, where each chunk has $b$ bits, and each precomputed table has $2^b$ elements. 
We call the table requirement LUT. Instead of building a single massive table, we create $ \lceil \frac{R^3}{b} \rceil $ requirement LUTs built from compression LUT as we prepare a table for all bit chunks. A list of masks out of the requirement LUTs can be combined into a bit sequence by bitwise AND operations, where the combined bit sequence indicates the list of masks that satisfy all of the bit requirements of an object mask as shown in Fig.~\ref{fig:lutsearch}. The object mask can be conservatively replaced by any mask in the list. 

Once the conservative mask list is obtained, the mask with the lowest set bit count in the list is the tightest. 
Instead of the linear search, we use the least-significant bit (LSB) location to find the tightest mask by sorting the masks in the compression LUT order by their number of set bits before the requirement LUTs are built. Algorithm~\ref{alg:encmask} shows our mask search algorithm.

\begin{algorithm}%[!t]
\caption{ Search the optimal object mask in compression LUT. \textbf{Inputs, Functions, and Notations:} \textit{mask}: An object mask to search for.
\textit{R}: The mask resolution.
\textit{b}: The number of bits in a chunk that corresponds to the requirement table.
\textit{requirementTables}: The precomputed requirement tables.
\textit{$LSB(x)$}: Calculate least significant bit of x.
$\land$: A bitwise AND.}
\label{alg:encmask}
\SetKwProg{Fn}{function}{}{end}
\SetKwFunction{indexOfOptimalMask}{indexOfOptimalMask}
\SetKwFunction{LSB}{LSB}
\Fn{\indexOfOptimalMask{ mask, R, b, requirementTables }}
{
    {$nBatch$ $\gets$ $ \lceil \frac{R^3}{b} \rceil $ }\;
    \For{$i \gets 0$ to $nBatch - 1$} 
    {
        {$bits$ $\gets$ read $ b $ bits from $ mask $ at $ ( i \cdot b ) $-th bit }\;
        {$requirement_i$ $\gets$ $ requirementTables_{( i, bits)} $ }\;
    }

    {$conservativeList$ $\gets$ $requirement_0$ }\;
    \For{$i \gets 1 $ to $nBatch - 1$} 
    {
        $conservativeList$ $\gets$ $ conservativeList \land requirement_i $\;
    }
    \Return \LSB( $conservativeList$ )\;
}
\end{algorithm}

\subsubsection{BVH Node Structure}
\begin{figure}
    \centering
    \begin{subfigure}[b]{0.49\textwidth}
    \includegraphics[width=1\linewidth,trim=180 100 180 130, clip]{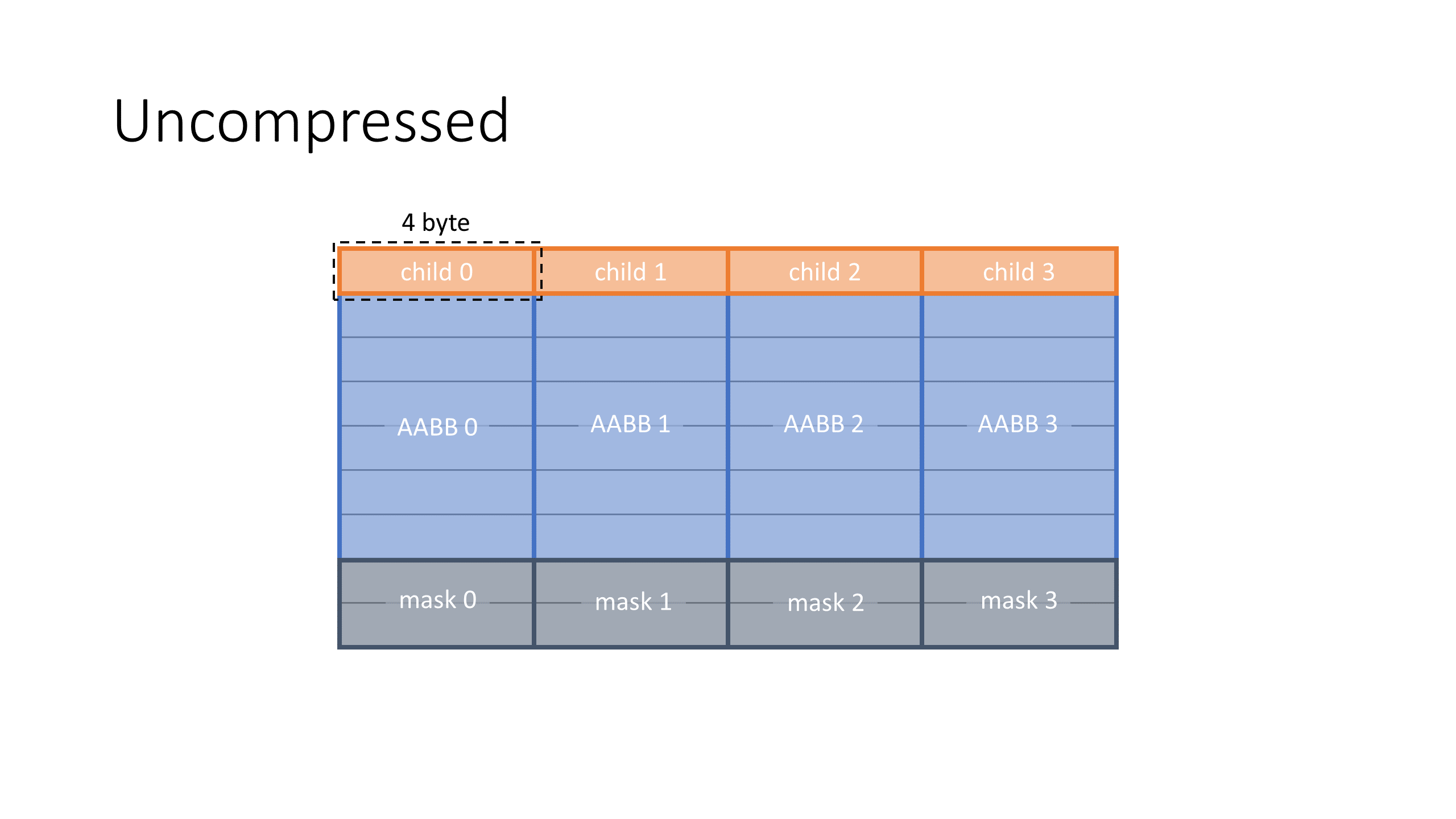} % trim: left, bottom, right and top
    \caption{ Uncompressed }
    \label{fig:uncompressed}
    \end{subfigure}
    \hfill
    \begin{subfigure}[b]{0.49\textwidth}
    \includegraphics[width=1\linewidth,trim=180 100 180 130, clip]{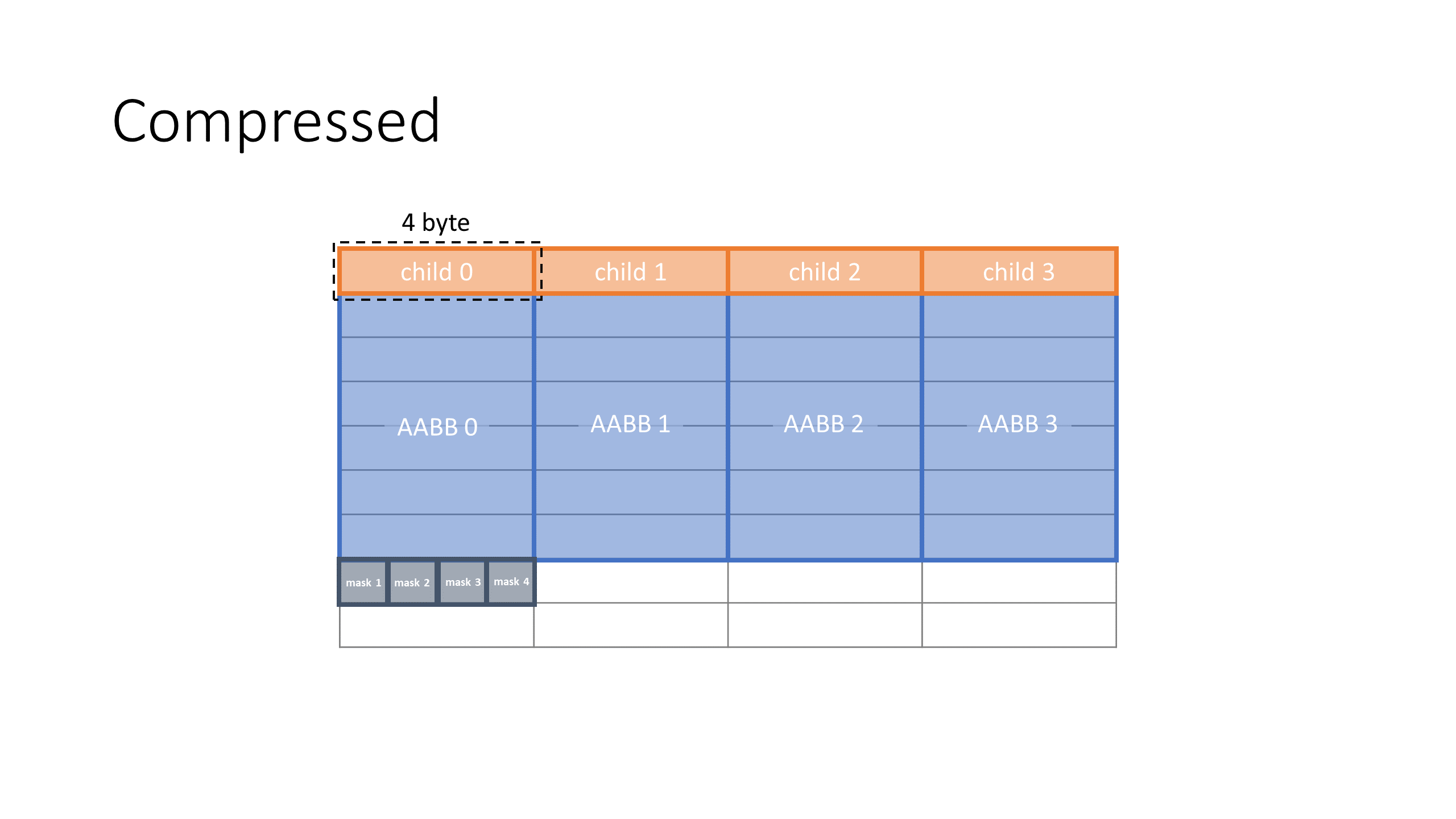} % trim: left, bottom, right and top
    \caption{ Compressed }
    \label{fig:compressed}
    \end{subfigure}
    \caption{Visualization of memory layout for 4-wide BVH nodes with our object masks, where $R=4$. Each rectangle cell indicates 4 bytes, and each square cell indicates 1 byte. (a): A BVH node without compression. (b): A BVH node with our compression by the LUT. }
    \label{fig:bvhlayout}
\end{figure}

% We store object masks for each BVH node alongside the bounding box. An example of our data structure is shown in Fig.~\ref{fig:bvh4}.
% When the number of the compression LUT element is 256, the index of a mask can be stored in 1 byte. 
Fig.~\ref{fig:bvhlayout} shows memory layout visualization of 4-wide BVH nodes with our object masks as examples, where $R=4$, the number of the compression LUT element is 256, Fig.~\ref{fig:uncompressed} is without the compression, and Fig.~\ref{fig:compressed} is with the compression. Since the number of bits in the object mask is $ R^3 $, each mask in the uncompressed node is 8 bytes. The index of a mask in the compression LUT can be stored in 1 byte, which is $\frac{1}{8}$ of the uncompressed mask. The size of the uncompressed node is 144 bytes compared to 112 bytes without masks. The size of a node with compression is 116 bytes.

\subsubsection{Traversal Algorithm}
Our subspace culling algorithm can be added to an existing regular BVH traversal algorithm. A pseudocode is shown in Algorithm~\ref{alg:bvhtraverse} where the addition of the logic for the proposed method is highlighted in red. Our method reuses the results of the AABB-ray intersection.

\begin{algorithm}%[!t]
\caption{ BVH traversal with our culling. The red highlighted part is our culling embedded in a traditional BVH traversal. \textbf{Inputs, Functions, and Notations:} \textit{root}: A root node of the BVH.
\textit{ray}: A ray to find intersections to the BVH.
\textit{R}: The mask resolution.
\textit{rayMasks}: The LUT for a ray mask. Discretized begin and end location of the ray map to its ray mask.
\textit{lookupRayMask}: LUT-based ray mask. Creation of the LUT is described in Algorithm~\ref{alg:lookup} }
\label{alg:bvhtraverse}
\SetKwProg{Fn}{function}{}{end}
\SetKwFunction{traverse}{traverse}
\SetKwFunction{push}{push}
\SetKwFunction{pop}{pop}
\SetKwFunction{lookupRayMask}{lookupRayMask}
\SetKwFor{While}{while}{}{end while}
\SetKw{Continue}{continue}
\Fn{\traverse{ root, ray, R, rayMasks }}
{
    \push($root$)\;
    \While{ $True$ }
    {
        $node$ $\gets$ \pop() \;
        \lIf{ $node$ is empty } { \textbf{break} }
        \If{ $node$ is a leaf } 
        { 
            Find an intersection $ray$ and triangles in $node$ \;
            \Continue\; 
        }
        {$hits$ $\gets$ find intersections AABBs in $node$ and $ray$ }\;
        \ForEach{ $hit_i \in hits$ }
        {
            \color{red}
            {
                $AABB$, $objectMask$, $ps$ $\gets$ get an AABB, an object mask, and intersections of $hit_i$\;
                $rayMask$ $\gets$ \lookupRayMask( $AABB_{lower}, AABB_{upper}, ps_0, ps_1, R, rayMasks$ )\;
                \If{ $objectMask \land rayMask $ is not zero } 
                {
                    \color{black}
                    {
                        $child$ $\gets$ get an child node that corresponds $hit_i$\;
                        \push($child$)\;
                    }
                }
            }
        }
    }
}
\end{algorithm}

\section{Evaluation}
\begin{figure}
    \begin{subfigure}[b]{0.325\textwidth}
        \centering
        \includegraphics[width=\textwidth]{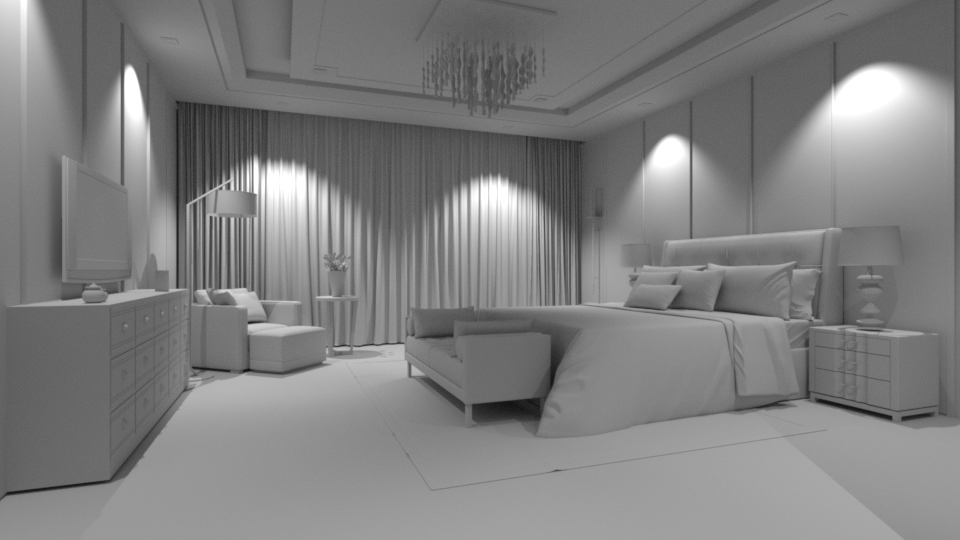}
        \caption{Bedroom (462.8 K tris)}
        \label{fig:sceneimages:a}
    \end{subfigure}
    \hfill
    \begin{subfigure}[b]{0.325\textwidth}
        \centering
        \includegraphics[width=\textwidth]{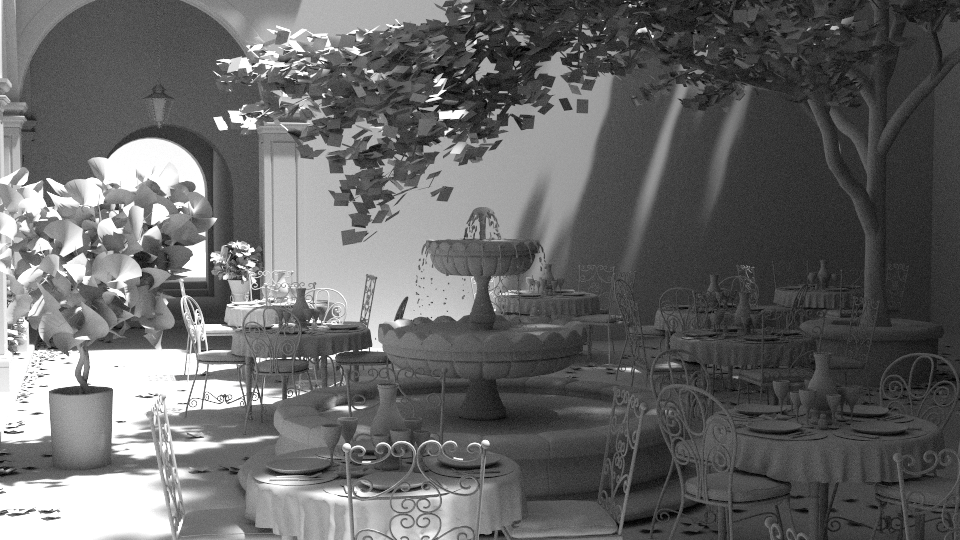}
        \caption{San Miguel (9.9 M tris)}
        \label{fig:sceneimages:b}
    \end{subfigure}
    \hfill
    \begin{subfigure}[b]{0.325\textwidth}
        \centering
        \includegraphics[width=\textwidth]{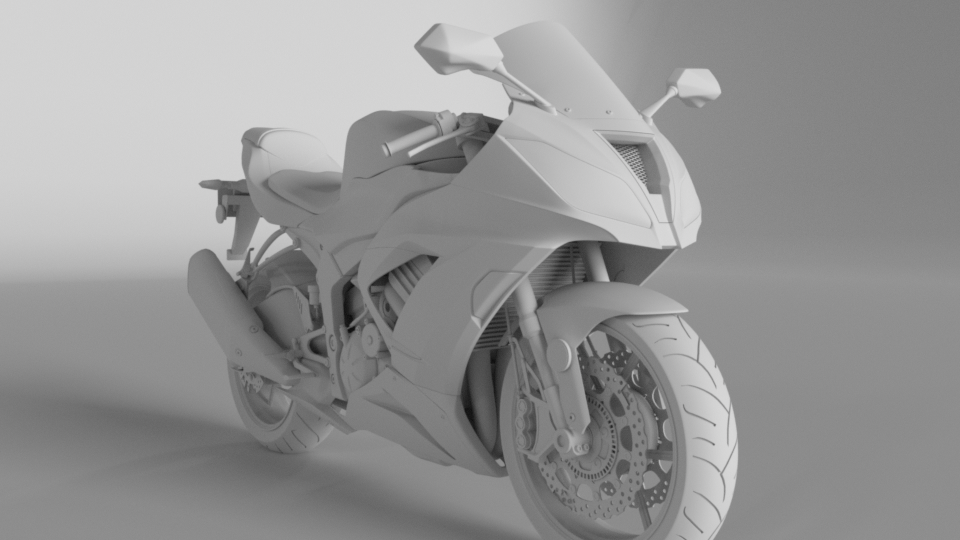}
        \caption{Ninja (1.3 M tris)}
        \label{fig:sceneimages:c}
    \end{subfigure}
    \hfill
    \begin{subfigure}[b]{0.325\textwidth}
        \centering
        \includegraphics[width=\textwidth]{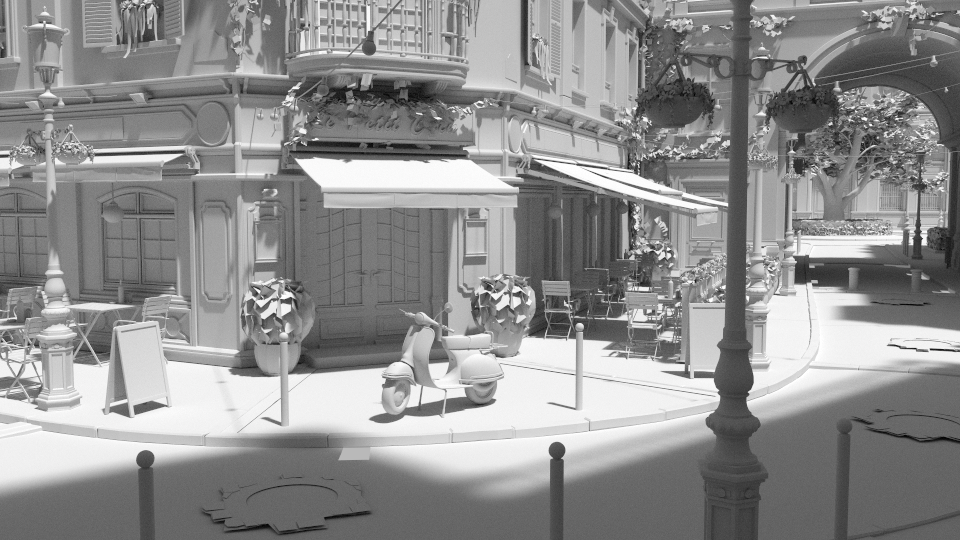}
        \caption{Bistro (2.8 M tris)}
        \label{fig:sceneimages:d}
    \end{subfigure}
    \hfill
    \begin{subfigure}[b]{0.325\textwidth}
        \centering
        \includegraphics[width=\textwidth]{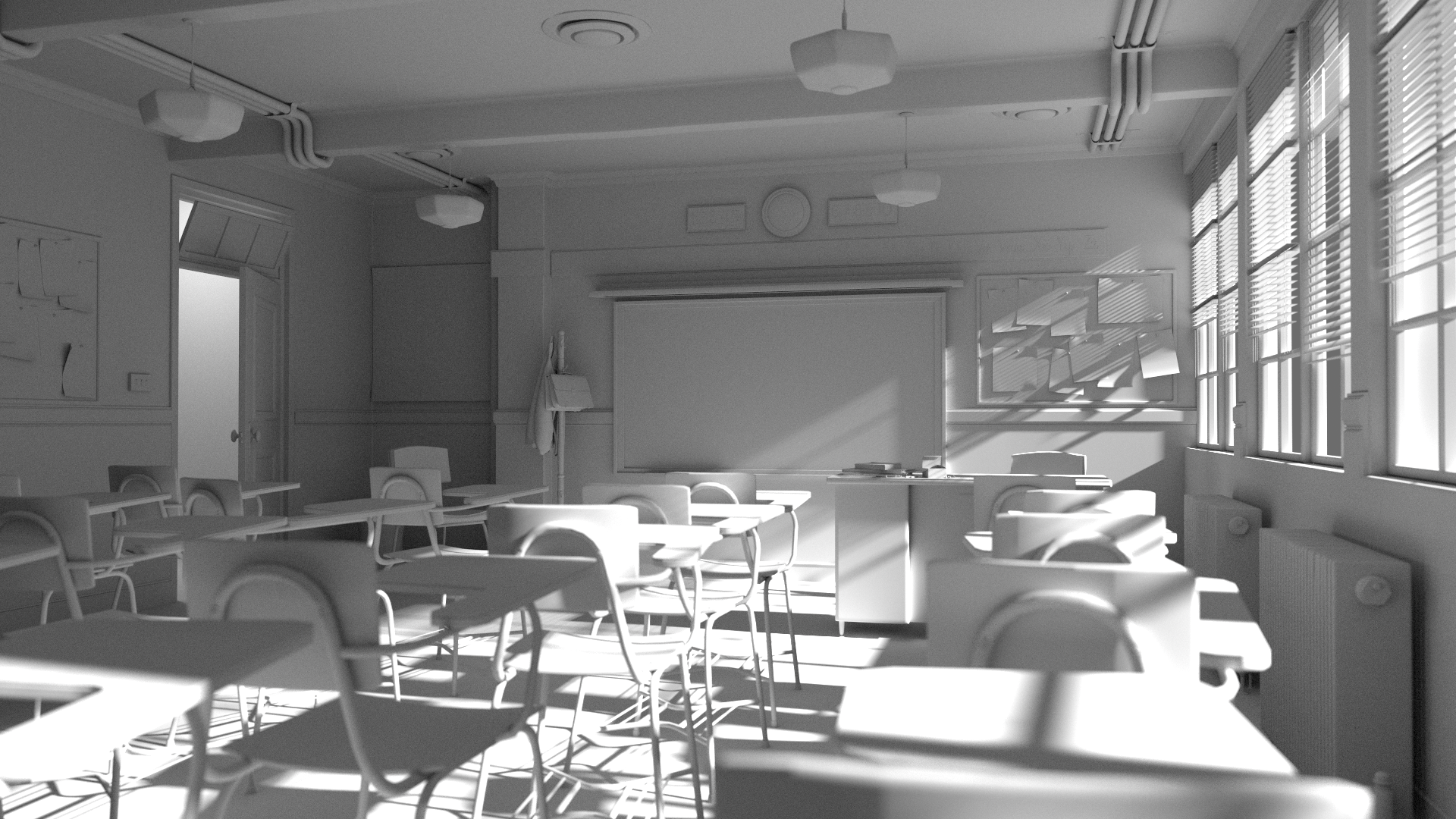}
        \caption{Classroom (606.1 K tris)}
        \label{fig:sceneimages:e}
    \end{subfigure}
    \hfill
    \begin{subfigure}[b]{0.325\textwidth}
        \centering
        \includegraphics[width=\textwidth]{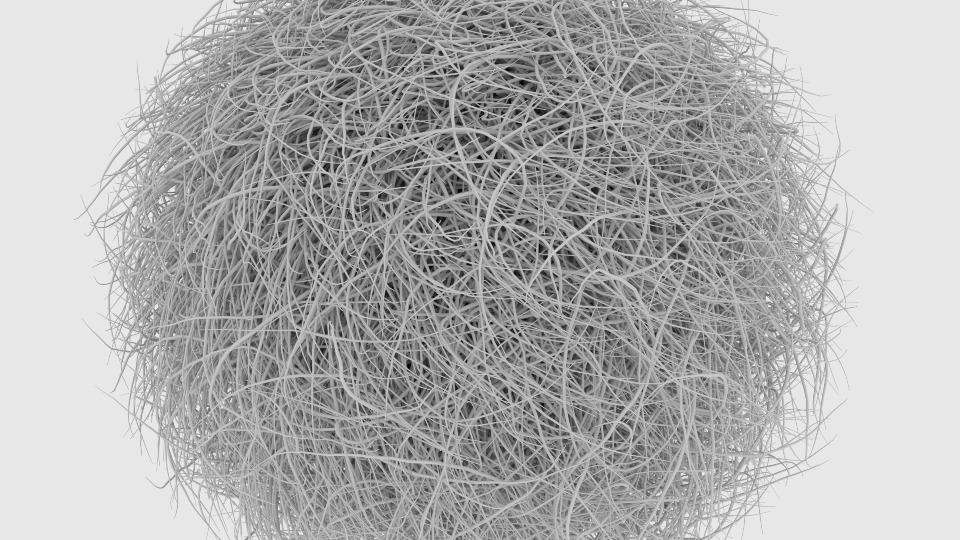}
        \caption{Hairball (2.8 M tris)}
        \label{fig:sceneimages:f}
    \end{subfigure}
    \hfill 
    \begin{subfigure}[b]{0.325\textwidth}
        \centering
        \includegraphics[width=\textwidth]{fig/scenes/CurlyHair.png}
        \caption{Curly Hair (12.1 M tris)}
        \label{fig:sceneimages:g}
    \end{subfigure}
    \hfill
    \begin{subfigure}[b]{0.325\textwidth}
        \centering
        \includegraphics[width=\textwidth]{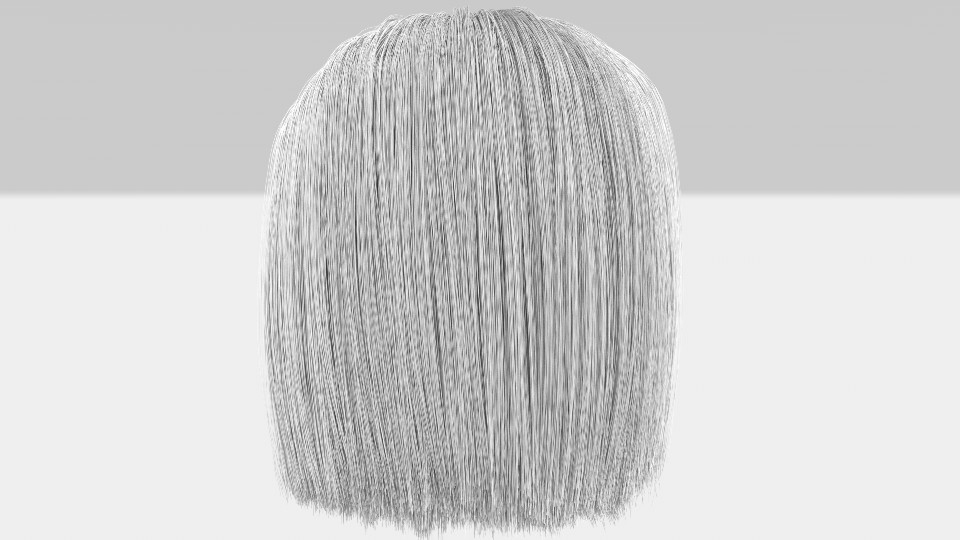}
        \caption{Straight Hair (7.3 M tris)}
        \label{fig:sceneimages:h}
    \end{subfigure}
    \hfill    
    \begin{subfigure}[b]{0.325\textwidth}
        \centering
        \includegraphics[width=\textwidth]{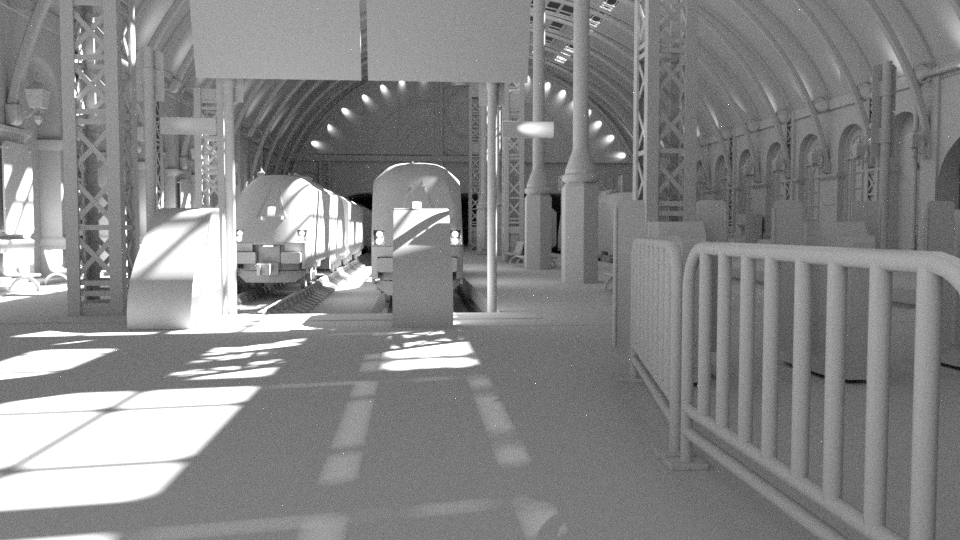}
        \caption{Victorian Trains (884.1 K tris)}
        \label{fig:sceneimages:i}
    \end{subfigure}
    \caption{ Test scenes used to evaluate our subspace culling. }
    \label{fig:sceneimages}
\end{figure}

We evaluate the culling efficiency in path tracing with several scenes shown in Fig.~\ref{fig:sceneimages}.
All the measurements are done with 960 $\times$ 540 resolution and 16 samples per pixel. % I3D'23 final acceptance notification for paper 18
We count the number of intersections of AABB-ray and triangle-ray to evaluate the culling efficiency.
As we use 4-wide BVH for all measurements in this paper, we treat four AABB-ray intersections and a triangle-ray intersection as a measurement unit. \revone{The BVH for each scene is built by binned SAH BVH construction algorithm \cite{BVHStar}.} Note that our method does not assume any branching factor or BVH construction algorithm; therefore, it applies to a BVH with any of them.
The maximum culling efficiency with our method is achieved by ray masks created with iterative grid traversal and object masks built without the approximation where an object mask is calculated by all of the descendant primitives in the BVH.
First, we show the maximum culling efficiency. Next, we evaluate the increase of intersections and their rendering performance with the lookup table optimization of the ray mask. The trade-off between the culling efficiency and the performance of object mask building with approximated object masks is also shown. 
Finally, we show the culling efficiency of the compressed masks and their object mask-finding performance. All measurements were done with an AMD Ryzen\textsuperscript{\texttrademark}9 5950X CPU. 

% better to combine 2 tables?
% https://www.embree.org/papers/2014-HPG-hair.pdf

\subsection{Ideal Culling}
We use grid traversal proposed by Amanatides et al. for the maximum culling efficiency \cite{fastvoxel}.
"Ideal Culling" column in Table~\ref{tbl:isects} shows the ratio of intersections compared to without our culling.
The ideal intersection ratio is 75.4\% on average, 56.5\% on the best with $R=4$, and 65.3\% on average, 42.5\% on the best with $R=6$. 
\revone{ Significant intersection reductions are observed in the "Curly Hair" and "Straight Hair" scene compared to the others. It emphasizes our method is suitable for thin and tilted geometry. }
Despite the reduction of intersections, it is difficult to pay off the culling overhead due to the ray mask creation by grid traversal as shown in Table~\ref{tbl:perf}.
% The maximum culling efficiency with 
% The tightest mask for AABB occupancy is obtained by querying all primitives in the volume. 
% Also, The tightest ray mask is built by grid traversal \cite{fastvoxel}. 
% The ideal culling ratio with our method is shown as "Ideal Culling" column in Table~\ref{tbl:isects}. 
% The culling ratio is 75.4\% on average with $R=4$, and 65.3\% on average with $R=6$. 
% The traversal performance of the ideal culling is also shown in Table~\ref{tbl:perf}. 
% It is difficult to pay off the culling overhead due to the ray mask creation by grid traversal despite the culling ratio being 75.4\% on average with $R=4$, 65.3\% on average with $R=6$. 
% make it positive -> negative flow

\subsection{Ray Mask Creation with LUT}
Although the LUT for ray masks can be used to avoid the expensive grid traversal, extra false positives are produced due to the use of the LUT. 
We evaluate the increase of intersections with ray mask LUT and the rendering performance. The LUT resolution--$R_{ray}$ can be independent of mask resolution--$R$. However, a LUT cell can overlap with more than one cell in an AABB mask if $R_{ray}$ is not multiple of $R$ or $R_{ray}$ is less than $R$. 
These misaligned grids reduce the method effectiveness because it produces false positives which could be culled.
% \atsushi{This discrepancy produces worse tightness of the ray mask. (There may be a better way to describe this. 1 or 2 lines)} 
% \takahiro{[todo. how about this? This reduces the effectiveness of the method because it produces false positives which could be culled if the grids were aligned.]}
% \atushi{ Looks better! }
% which makes the test operation complex. 
Thus, we use $R_{ray}=R$ and $R_{ray}=2R$ for the evaluation. 
The creation of ray mask LUT with $R=4$ and $R=6$ take 1.1 milliseconds and 29.6 milliseconds, respectively, if $R_{ray}=R$, while they take 54.2 milliseconds and 1791.7 milliseconds if $R_{ray}=2R$.
The two intersection ratios and performance measurements are also shown in Table~\ref{tbl:isects} and Table~\ref{tbl:perf}, respectively.
Although the intersection ratio of ray mask LUT $R_{ray}=R, 2R$ is slightly worse than the ideal case, all scenes get faster than the ideal culling.
Especially, over 10\% render time reductions from the baseline are observed in "Curly Hair" and "Straight Hair" scenes with $R_{ray}=R$. 
$R_{ray}=2R$ case is slower than $R_{ray}=R$ despite of the better tightness. This can be explained latency penalty of the memory access to random locations in the larger size of the LUT.

% just \textbf looked not that standing out
\newcommand\M[1]{\underline{\textbf{#1}}}

\begin{table}
\small
\begin{tabular}{ |p{2.1cm}||p{1.4cm}|p{2.2cm}|p{2.2cm}|p{2.2cm}|p{2.2cm}|  }
 \hline
 \multicolumn{6}{|c|}{ Ratio of intersections with resolution $R=4$ } \\
 \hline
 Scene & No Culling & Ideal Culling & Ray mask LUT $R_{ray}=R$ & Ray mask LUT $R_{ray}=2R$ & Compressed with LUT \\
 \hline
Bedroom & 100\% &  89.3\% & 90.0\% & 89.8\% & 90.7\% \\
San Miguel & 100\% &  79.3\% & 80.5\% & 80.2\% & 83.2\% \\
Ninja & 100\% &  84.2\% & 85.5\% & 85.2\% & 90.0\% \\
Bistro & 100\% &  72.3\% & 73.8\% & 73.4\% & 78.2\% \\
Classroom & 100\% &  87.4\% & 88.1\% & 87.9\% & 88.3\% \\
Hairball & 100\% &  67.0\% & 71.5\% & 70.6\% & 73.2\% \\
Curly Hair & 100\% &  57.5\% & 62.5\% & 61.5\% & 66.3\% \\
Straight Hair & 100\% &  \M{56.5\%} & \M{61.4\%} & \M{60.3\%} & \M{62.1\%} \\
VictorianTrains & 100\% &  85.1\% & 86.2\% & 85.9\% & 86.9\% \\
\hline
\textit{Average} & 100\% &  75.4\% & 77.7\% & 77.2\% & 79.9\% \\
\hline
 \multicolumn{6}{|c|}{ Ratio of intersections with resolution $R=6$ } \\
 \hline
Bedroom & 100\% &  83.8\% & 84.9\% & 84.5\% & 86.9\% \\
San Miguel & 100\% &  67.9\% & 70.6\% & 69.7\% & 76.5\% \\
Ninja & 100\% &  76.5\% & 78.8\% & 78.0\% & 87.7\% \\
Bistro & 100\% &  59.1\% & 61.8\% & 61.0\% & 71.5\% \\
Classroom & 100\% &  83.5\% & 83.3\% & 83.7\% & 85.1\% \\
Hairball & 100\% &  54.4\% & 61.1\% & 58.8\% & 74.3\% \\
Curly Hair & 100\% &  43.3\% & 50.6\% & 48.1\% & 60.4\% \\
Straight Hair & 100\% &  \M{42.5\%} & \M{49.3\%} & \M{47.0\%} & \M{53.6\%} \\
VictorianTrains & 100\% &  77.0\% & 78.7\% & 78.1\% & 81.1\% \\
\hline
\textit{Average} & 100\% &  65.3\% & 68.8\% & 67.6\% & 75.2\% \\
\hline
\end{tabular}
\caption{ Ratio of AABB-Ray and AABB-Triangle intersections with and without our culling. Ideal culling case, ray mask with LUT approximation, and compressed cases are shown. The underlined numbers are the best ratio in the scenes. }
\label{tbl:isects}
\end{table}

\begin{table}
\small
\begin{tabular}{ |p{2.1cm}||p{3.0cm}|p{2.3cm}|p{2.3cm}|p{2.3cm}|  }
 \hline
 \multicolumn{5}{|c|}{ Relative rendering time with resolution $R=4$ } \\
 \hline
 Scene & No Culling & Ideal Culling & Ray mask LUT $R_{ray}=R$ & Ray mask LUT $R_{ray}=2R$ \\
 \hline
Bedroom & 100\% ( 73.89 seconds ) &  117.2\% & 103.6\% & 103.0\% \\
San Miguel & 100\% ( 180.03 seconds ) &  138.0\% & 97.6\% & 99.3\% \\
Ninja & 100\% ( 55.19 seconds ) &  \M{113.7}\% & 101.0\% & 101.2\% \\
Bistro & 100\% ( 97.01 seconds ) &  124.8\% & 94.9\% & 97.1\% \\
Classroom & 100\% ( 123.31 seconds ) &  120.2\% & 106.0\% & 102.9\% \\
Hairball & 100\% ( 54.11 seconds ) &  132.1\% & 97.9\% & 98.0\% \\
Curly Hair & 100\% ( 82.74 seconds ) &  126.7\% & \M{86.9}\% & \M{89.2}\% \\
Straight Hair & 100\% ( 54.34 seconds ) &  120.0\% & 88.0\% & 91.1\% \\
VictorianTrains & 100\% ( 129.19 seconds ) &  124.4\% & 104.3\% & 107.2\% \\
% Bedroom & 100\% &  117.2\% & 103.6\% & 103.0\% \\
% San Miguel & 100\% &  138.0\% & 97.6\% & 99.3\% \\
% Ninja & 100\% &  \M{113.7\%} & 101.0\% & 101.2\% \\
% Bistro & 100\% &  124.8\% & 94.9\% & 97.1\% \\
% Classroom & 100\% &  120.2\% & 106.0\% & 102.9\% \\
% Hairball & 100\% &  132.1\% & 97.9\% & 98.0\% \\
% Curly Hair & 100\% &  126.7\% & \M{86.9\%} & \M{89.2\%} \\
% Straight Hair & 100\% &  120.0\% & 88.0\% & 91.1\% \\
% VictorianTrains & 100\% &  124.4\% & 104.3\% & 107.2\% \\
\hline
\textit{Average} & 100\% ( 94.42 seconds ) &  124.1\% & 97.8\% & 98.8\% \\
% \textit{Average} & 100\% &  124.1\% & 97.8\% & 98.8\% \\
\hline
 \multicolumn{5}{|c|}{ Relative rendering time with resolution $R=6$ } \\
 \hline
Bedroom & 100\% ( 73.89 seconds ) &  119.7\% & 106.5\% & 107.9\% \\
San Miguel & 100\% ( 180.03 seconds ) &  140.0\% & 98.2\% & 108.4\% \\
Ninja & 100\% ( 55.19 seconds ) &  118.1\% & 104.5\% & 110.3\% \\
Bistro & 100\% ( 97.01 seconds ) &  127.1\% & 97.3\% & 105.7\% \\
Classroom & 100\% ( 123.31 seconds ) &  124.6\% & 107.1\% & 110.8\% \\
Hairball & 100\% ( 54.11 seconds ) &  132.5\% & 102.0\% & 110.9\% \\
Curly Hair & 100\% ( 82.74 seconds ) &  120.2\% & 87.1\% & 95.7\% \\
Straight Hair & 100\% ( 54.34 seconds ) &  \M{114.6}\% & \M{85.6}\% & \M{90.5}\% \\
VictorianTrains & 100\% ( 129.19 seconds ) &  128.4\% & 110.0\% & 115.8\% \\
% Bedroom & 100\% &  119.7\% & 106.5\% & 107.9\% \\
% San Miguel & 100\% &  140.0\% & 98.2\% & 108.4\% \\
% Ninja & 100\% &  118.1\% & 104.5\% & 110.3\% \\
% Bistro & 100\% &  127.1\% & 97.3\% & 105.7\% \\
% Classroom & 100\% &  124.6\% & 107.1\% & 110.8\% \\
% Hairball & 100\% &  132.5\% & 102.0\% & 110.9\% \\
% Curly Hair & 100\% &  120.2\% & 87.1\% & 95.7\% \\
% Straight Hair & 100\% &  \M{114.6\%} & \M{85.6\%} & \M{90.5\%} \\
% VictorianTrains & 100\% &  128.4\% & 110.0\% & 115.8\% \\
\hline
\textit{Average} & 100\% ( 94.42 seconds ) &  125.0\% & 99.8\% & 106.2\% \\
% \textit{Average} & 100\% &  125.0\% & 99.8\% & 106.2\% \\
\hline
\end{tabular}
\caption{ Rendering time comparison for ray mask LUT. The ratios are relative rendering times with respect to without culling. The underlined numbers are the best performance in the scenes. }
\label{tbl:perf}
\end{table}

\subsection{Approximated Object Mask}
\begin{figure}
\begin{subfigure}[b]{0.49\textwidth}
    \includegraphics[width=1\linewidth]{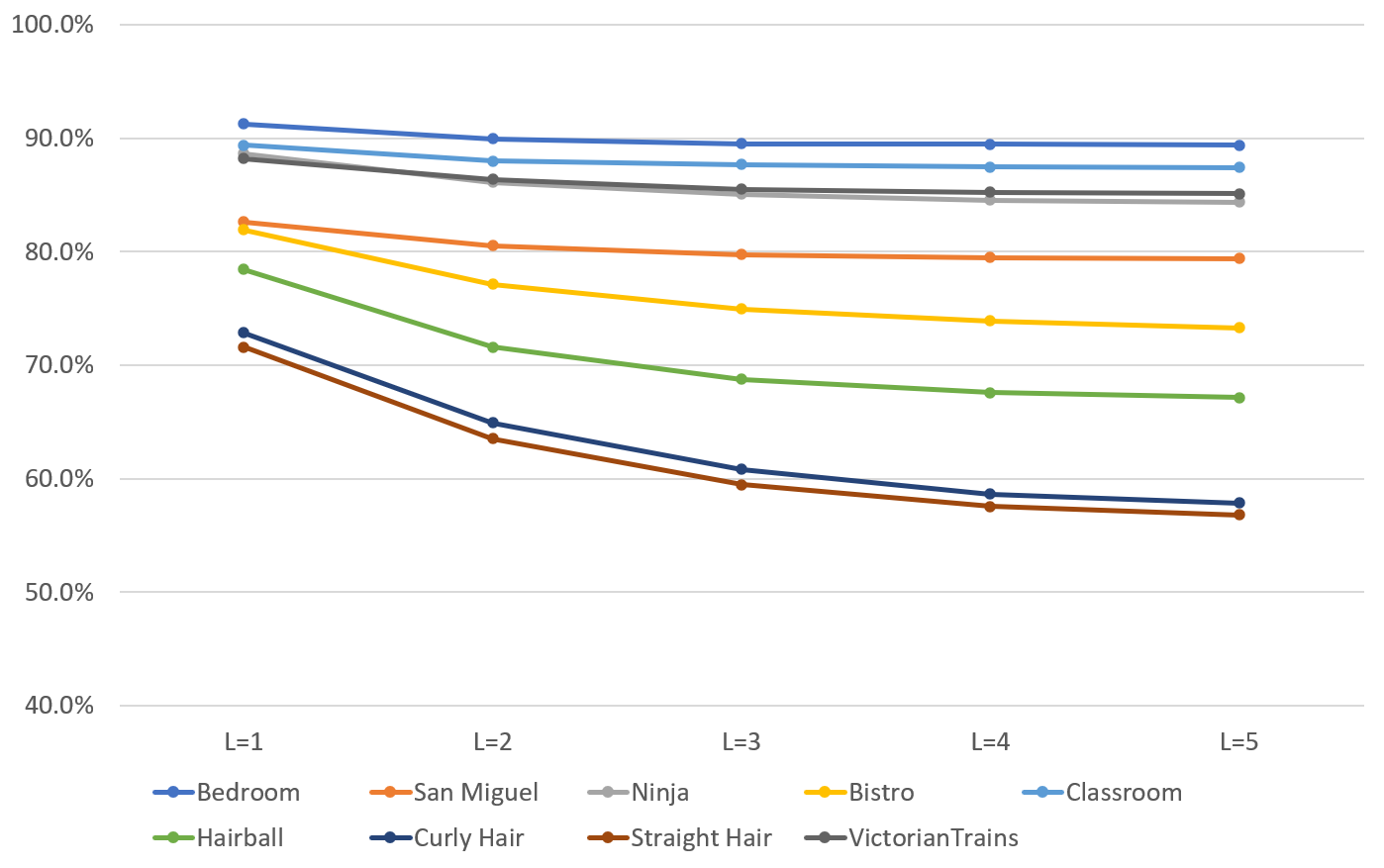} 
    \caption{ \revone{Ratio of intersections} with $R=4$ }
\end{subfigure}
\begin{subfigure}[b]{0.49\textwidth}
    \includegraphics[width=1\linewidth]{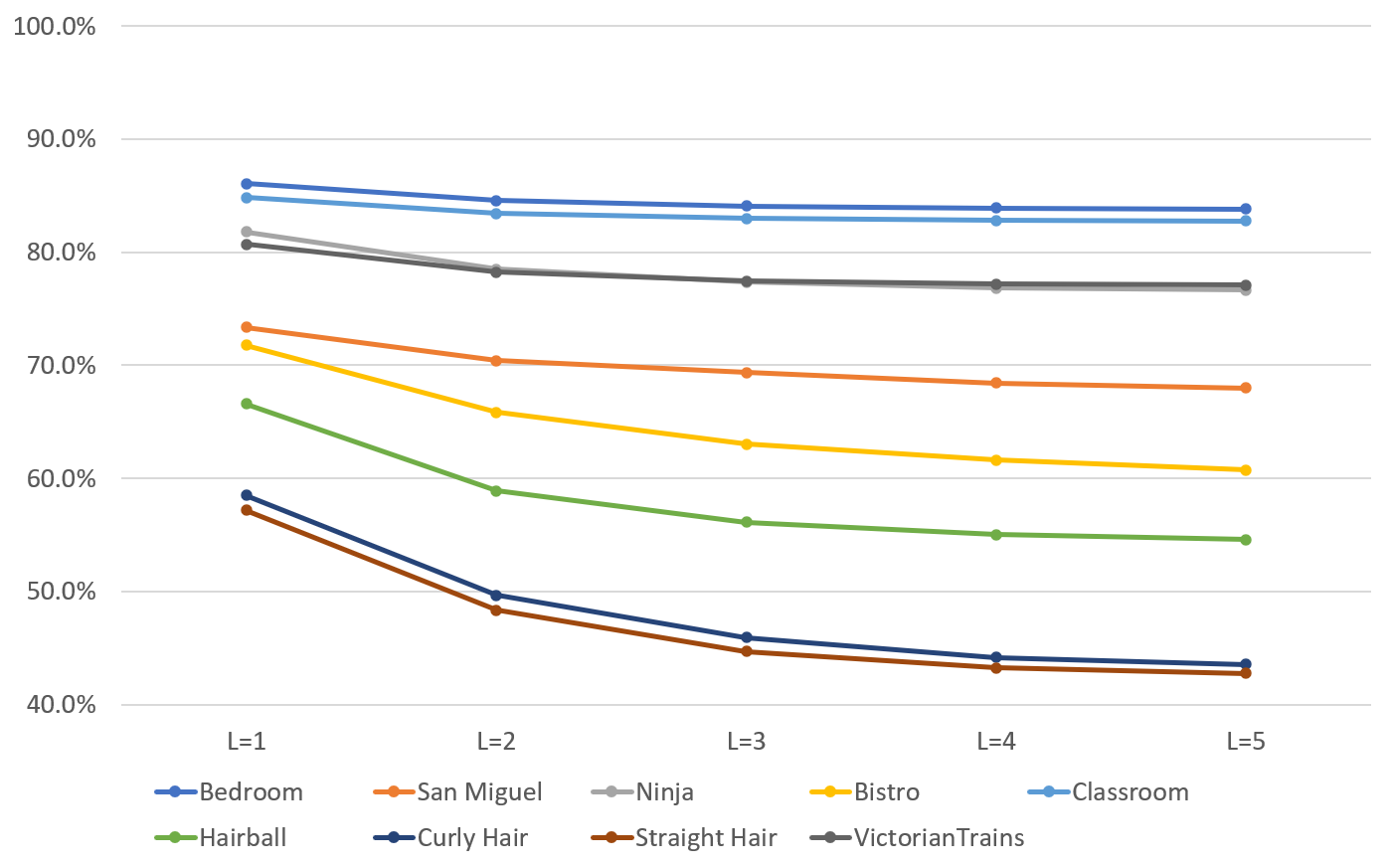} 
    \caption{ \revone{Ratio of intersections} with $R=6$  }
\end{subfigure}
\begin{subfigure}[b]{0.49\textwidth}
    \includegraphics[width=1\linewidth]{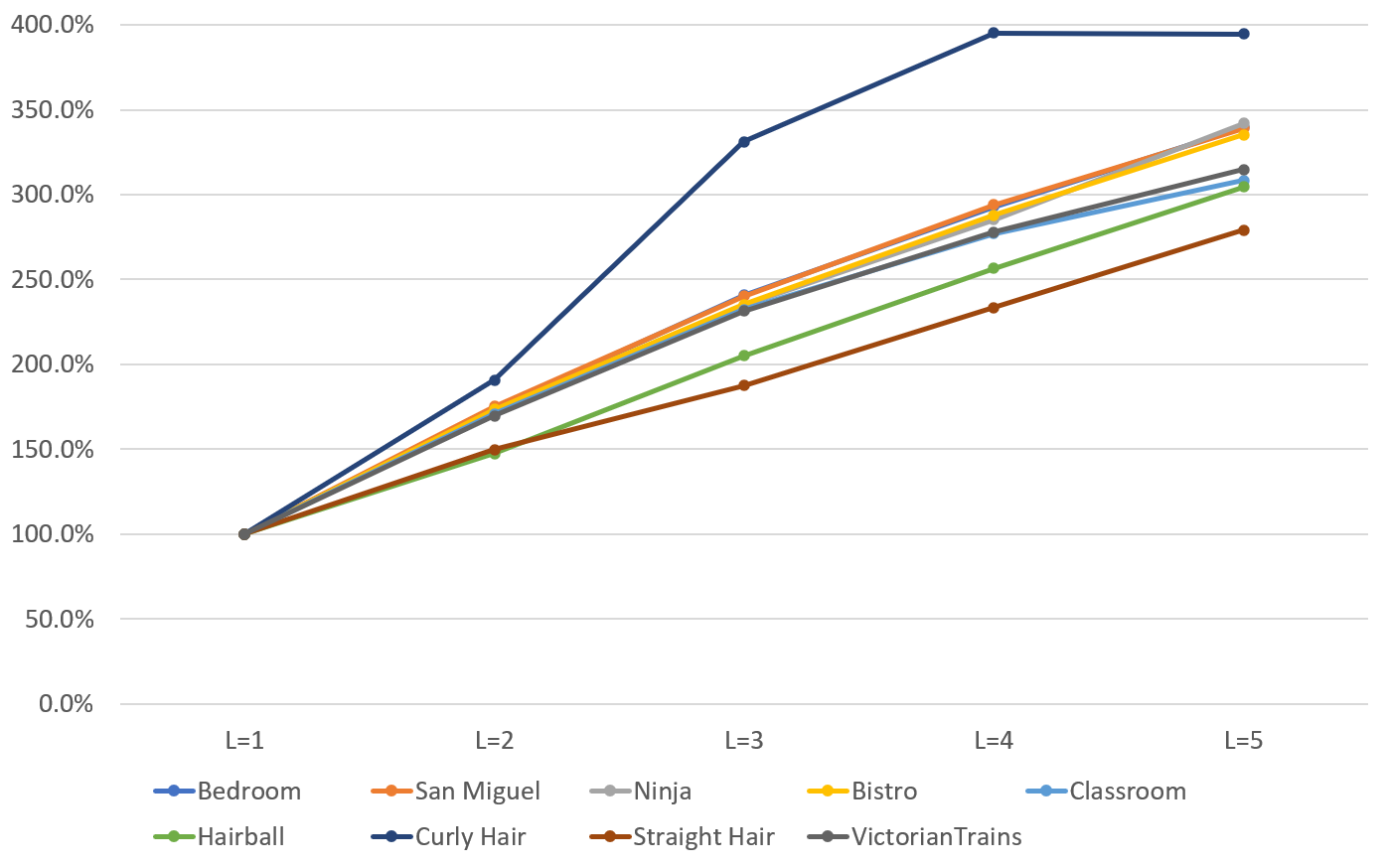} 
    \caption{ Object mask creation time with $R=4$ }
\end{subfigure}
\begin{subfigure}[b]{0.49\textwidth}
    \includegraphics[width=1\linewidth]{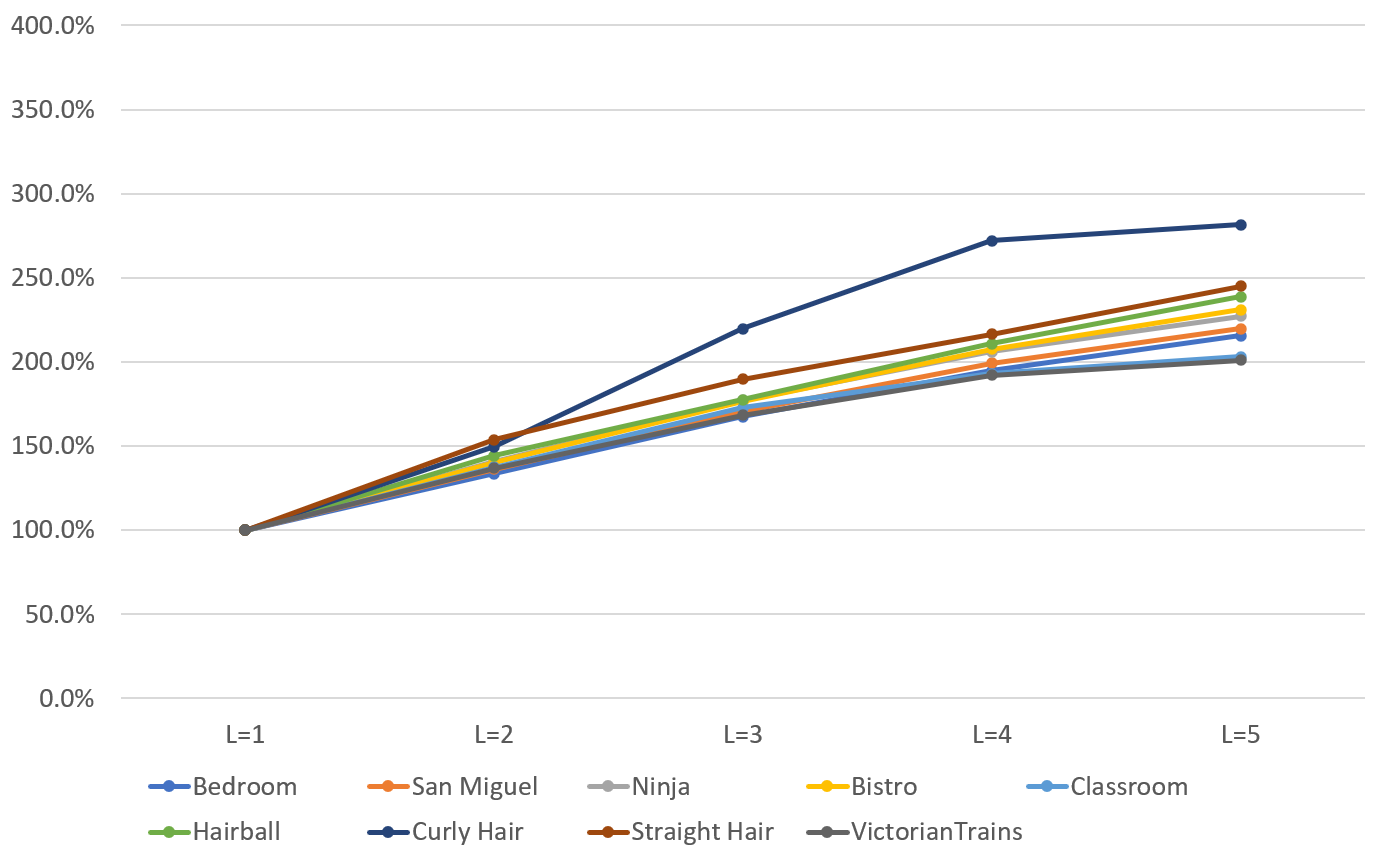} 
    \caption{ Object mask creation time with $R=6$  }
\end{subfigure}
    
    \caption{ (a) and (b): \revone{Ratio of intersections} with object mask approximation with parameter $L=1,2,3,4,5$ and the resolution $R=4,6$. (c) and (d): Object mask relative creation time with the approximation with parameter $L=1,2,3,4,5$ and the resolution $R=4,6$. The baseline of the ratio is $L=1$, }
    \label{fig:cullingchart}
\end{figure}

We evaluate the approximation of the object mask construction in a BVH with different parameters where $L= 1, 2, 3, 4, 5$.
As the tightness and its computational cost depend on the parameter $L$, we measure the intersection and the creation time of the mask.
Fig.~\ref{fig:cullingchart} shows the trade-off between the intersection ratio and the mask creation performance. However, proper $L$ depends on the performance gain from the reduction and the amount of ray-tracing cost. As the reduction of intersection starts saturated around $L=3$ while the mask creation time increase near to linear with respect to $L$ on almost all scenes, $L=3$ could be used as a default parameter before fine-tuning.

\subsection{Compressed Object Mask}
As the bit pattern of the object mask can be alternated by another object mask as long as it is conservative, the number of the LUT elements can be arbitrary.
We evaluate the compression with 256 for the LUT size for simplicity of the compression and decompression.
The compression rate is fixed, which is 1:8 for $R=4$ and 1:27 for $R=6$.
Table~\ref{tbl:isects} shows the intersection ratio as "Compressed with LUT" column. Despite the compressed size being the same, $R=6$ case produces better culling ratios with compression. 24.8\% on average, 46.4\% on maximum was reduced by just 8bits mask data per AABB in $R=6$ case.

We also evaluate the performance of the object mask finding. We use $b=8$ for mask search for efficient bit extraction from the target object mask and keeping the table size small.
Table~\ref{tbl:masksearch} shows a performance comparison between the brute-force search and our search. The latter only takes 9.1\%—11.1\% with $R=4$, 10.7\%—17.2\% with $R=6$, compared with the former.

\begin{table}
\small
\begin{tabular}{ |p{2.1cm}||p{2.7cm}|p{2.7cm}|p{2.1cm}|  }
 \hline
 \multicolumn{4}{|c|}{ Mask finding time $R=4$ } \\
 \hline
 Scene & Naive (milliseconds) & Our (milliseconds) & Our (Relative) \\
 \hline
Bedroom &  216.6 & 20.3 &9.4\% \\
San Miguel &  4923.2 & 446.6 &9.1\% \\
Ninja &  569.6 & 63.1 &11.1\% \\
Bistro &  1295.2 & 128.5 &9.9\% \\
Classroom &  279.7 & 26.4 &9.4\% \\
Hairball &  1396.5 & 132.2 &9.5\% \\
Curly Hair &  5325.4 & 543.0 &10.2\% \\
Straight Hair &  3132.5 & 326.7 &10.4\% \\
VictorianTrains &  409.8 & 37.4 &9.1\% \\
\hline
 \multicolumn{4}{|c|}{ Mask finding time $R=6$} \\
 \hline
Bedroom &  371.2 & 46.8 &12.6\% \\
San Miguel &  9181.5 & 983.7 &10.7\% \\
Ninja &  1239.8 & 139.1 &11.2\% \\
Bistro &  2646.1 & 305.9 &11.6\% \\
Classroom &  515.4 & 60.9 &11.8\% \\
Hairball &  2039.0 & 306.8 &15.0\% \\
Curly Hair &  8768.3 & 1251.3 &14.3\% \\
Straight Hair &  4522.9 & 777.2 &17.2\% \\
VictorianTrains &  674.5 & 95.0 &14.1\% \\
\hline
\end{tabular}
\caption{ Time of searching the conservative and tightest object mask in a table for compression. }
\label{tbl:masksearch}
\end{table}

% \begin{figure}
%   % \centering
%   % \includegraphics[width=6.0in]{fig/teaser.png}
%   % \caption{Drumheller Fountain, The University of Washington, Seattle WA.}
%   \begin{subfigure}[b]{0.49\textwidth}
%     \begin{overpic}[width=1\linewidth]{fig/measure/R4_Perf.png} 
%     \end{overpic}
%     \caption{ $R=4$ }
%     \label{teaser:without}
%     \end{subfigure}
%     \begin{subfigure}[b]{0.49\textwidth}
%     \begin{overpic}[width=1\linewidth]{fig/measure/R6_Perf.png} 
%     \end{overpic}
%     \caption{ $R=6$  }
%     \label{teaser:with}
%     \end{subfigure}
%     \caption{ Culling rate with different parameters }
%     \label{perfchart}
% \end{figure}
\section{ Conclusions and Future work }
We proposed a novel culling technique using ray and object masks that can be embedded in a BVH node. 
Our ray LUT-based approach reduces the number of intersections by 38.6\% with $R=4$ and 50.7\% with $R=6$ as the maximums in our experiment.
Additionally, the traversal performance with the ray LUT approach shows performance improvement in certain scenes where AABB causes too many false positive intersections due to the loose fit.
Approximated object mask provides a trade-off for object mask construction cost and culling efficiency. We also proposed compression of object masks using LUT without sacrificing much culling efficiency. 

Despite our algorithm's simplicity and the intersection reduction, it is difficult to achieve a better performance in all scenes, as we showed in Sec. 4, even though our computational overhead is small. A dedicated hardware implementation may change the algorithm trade-off between computational cost and culling efficiency. 
Optimal LUT creation for object mask compression in a practical time has yet to be established.
Fast and optimal compression LUT generation is important for the coexistence of culling efficiency, cheap computation, and low memory cost.

As each object mask in a BVH is affected by all primitives in their children's primitives, animated primitives enforce updating parent object masks. Efficient update approaches in dynamic scenes are left for future work. Another future work is an efficient object mask creation on the GPU and performance evaluation of the proposed method on the GPU.

% \takahiro{ [is this right?] > yes and new one looks nice.
% commented out for the first submission for the blind review.
\section{ acknowledgment }
\revone{We are grateful for all the data sets for our measurements and evaluation. 
"San Miguel" is by Guillermo M. Leal Llaguno. "Hairball" is from NVIDIA Research. 
"Bistro" is by Amazon Lumberyard. 
"Classroom" is by Christophe Seux from Blender Foundation's demo files. 
"Curly Hair" and "Straight Hair" are from Cem Yuksel's web page. 
We also thank Oleksandr Kupriyanchuk for proofreading and valuable suggestions. Additionally, we appreciate Prashanth Kannan for finding out a lot of grammatical mistakes and inconsistencies. }

% "Bedroom" is ours, don't have to mention it

\bibliographystyle{ACM-Reference-Format}
\bibliography{main}
\end{document}